%Paper: hep-ph/9303298
%From: Ulf Meissner <MEISSNER@ITP.unibe.ch>
%Date: Wed, 24 Mar 1993 10:16:54 +0100

% This is MATHMULT.DOC the documentation file of
% the plain TeX macro package from Springer-Verlag
% for multi author books in mathematics, version of March 1991
%\input mathmult.cmm
% This is MATHMULT.CMM the plain TeX macro package
% (CM version) from Springer-Verlag
% for multi author books in mathematics, version of March 1991
\font \tbfontt                = cmbx10 scaled\magstep1
\font \tafontt                = cmbx10 scaled\magstep2
\font \tbfontss               = cmbx5  scaled\magstep1
\font \tafontss               = cmbx5  scaled\magstep2
\font \sixbf                  = cmbx6
\font \tbfonts                = cmbx7  scaled\magstep1
\font \tafonts                = cmbx7  scaled\magstep2
\font \ninebf                 = cmbx9
\font \tasys                  = cmex10 scaled\magstep1

\font \sixi                   = cmmi6
\font \ninei                  = cmmi9
\font \tams                   = cmmib10
\font \tbmss                  = cmmib10 scaled 600
\font \tamss                  = cmmib10 scaled 700
\font \tbms                   = cmmib10 scaled 833
\font \tbmt                   = cmmib10 scaled\magstep1
\font \tamt                   = cmmib10 scaled\magstep2
\font \smallescriptscriptfont = cmr5
\font \smalletextfont         = cmr5 at 10pt
\font \smallescriptfont       = cmr5 at 7pt
\font \sixrm                  = cmr6
\font \ninerm                 = cmr9
\font \ninesl                 = cmsl9
\font \tensans                = cmss10
\font \fivesans               = cmss10 at 5pt
\font \sixsans                = cmss10 at 6pt
\font \sevensans              = cmss10 at 7pt
\font \ninesans               = cmss10 at 9pt
\font \tbst                   = cmsy10 scaled\magstep1
\font \tast                   = cmsy10 scaled\magstep2
\font \tbsss                  = cmsy5  scaled\magstep1
\font \tasss                  = cmsy5  scaled\magstep2
\font \sixsy                  = cmsy6
\font \tbss                   = cmsy7  scaled\magstep1
\font \tass                   = cmsy7  scaled\magstep2
\font \ninesy                 = cmsy9
\font \markfont               = cmti10 at 11pt
\font \nineit                 = cmti9
\font \ninett                 = cmtt9
%-----------------------------------------------------------------------
\magnification=\magstep0
%\hsize=12.2truecm
\hsize=13truecm
%\vsize=19.4truecm
\vsize=19.8truecm
\hfuzz=2pt
\tolerance=500
\abovedisplayskip=3 mm plus6pt minus 4pt
\belowdisplayskip=3 mm plus6pt minus 4pt
\abovedisplayshortskip=0mm plus6pt minus 2pt
\belowdisplayshortskip=2 mm plus4pt minus 4pt
\predisplaypenalty=0
\clubpenalty=10000
\widowpenalty=10000
\frenchspacing
\newdimen\oldparindent\oldparindent=1.5em
\parindent=1.5em
%-----------------------------------------------------------------------

 %reelle Zahlen

 %natuerliche Zahlen

\def\bbbc{{\mathchoice {\setbox0=\hbox{$\displaystyle\rm C$}\hbox{\hbox
to0pt{\kern0.4\wd0\vrule height0.9\ht0\hss}\box0}}
{\setbox0=\hbox{$\textstyle\rm C$}\hbox{\hbox
to0pt{\kern0.4\wd0\vrule height0.9\ht0\hss}\box0}}
{\setbox0=\hbox{$\scriptstyle\rm C$}\hbox{\hbox
to0pt{\kern0.4\wd0\vrule height0.9\ht0\hss}\box0}}
{\setbox0=\hbox{$\scriptscriptstyle\rm C$}\hbox{\hbox
to0pt{\kern0.4\wd0\vrule height0.9\ht0\hss}\box0}}}}
\def\bbbe{{\mathchoice {\setbox0=\hbox{\smalletextfont e}\hbox{\raise
0.1\ht0\hbox to0pt{\kern0.4\wd0\vrule width0.3pt
height0.7\ht0\hss}\box0}}
{\setbox0=\hbox{\smalletextfont e}\hbox{\raise 0.1\ht0\hbox
to0pt{\kern0.4\wd0\vrule width0.3pt height0.7\ht0\hss}\box0}}
{\setbox0=\hbox{\smallescriptfont e}\hbox{\raise 0.1\ht0\hbox
to0pt{\kern0.5\wd0\vrule width0.2pt height0.7\ht0\hss}\box0}}
{\setbox0=\hbox{\smallescriptscriptfont e}\hbox{\raise
0.1\ht0\hbox to0pt{\kern0.4\wd0\vrule width0.2pt
height0.7\ht0\hss}\box0}}}}
\def\bbbq{{\mathchoice {\setbox0=\hbox{$\displaystyle\rm Q$}\hbox{\raise
0.15\ht0\hbox to0pt{\kern0.4\wd0\vrule height0.8\ht0\hss}\box0}}
{\setbox0=\hbox{$\textstyle\rm Q$}\hbox{\raise
0.15\ht0\hbox to0pt{\kern0.4\wd0\vrule height0.8\ht0\hss}\box0}}
{\setbox0=\hbox{$\scriptstyle\rm Q$}\hbox{\raise
0.15\ht0\hbox to0pt{\kern0.4\wd0\vrule height0.7\ht0\hss}\box0}}
{\setbox0=\hbox{$\scriptscriptstyle\rm Q$}\hbox{\raise
0.15\ht0\hbox to0pt{\kern0.4\wd0\vrule height0.7\ht0\hss}\box0}}}}
\def\bbbt{{\mathchoice {\setbox0=\hbox{$\displaystyle\rm
T$}\hbox{\hbox to0pt{\kern0.3\wd0\vrule height0.9\ht0\hss}\box0}}
{\setbox0=\hbox{$\textstyle\rm T$}\hbox{\hbox
to0pt{\kern0.3\wd0\vrule height0.9\ht0\hss}\box0}}
{\setbox0=\hbox{$\scriptstyle\rm T$}\hbox{\hbox
to0pt{\kern0.3\wd0\vrule height0.9\ht0\hss}\box0}}
{\setbox0=\hbox{$\scriptscriptstyle\rm T$}\hbox{\hbox
to0pt{\kern0.3\wd0\vrule height0.9\ht0\hss}\box0}}}}
\def\bbbs{{\mathchoice
{\setbox0=\hbox{$\displaystyle     \rm S$}\hbox{\raise0.5\ht0\hbox
to0pt{\kern0.35\wd0\vrule height0.45\ht0\hss}\hbox
to0pt{\kern0.55\wd0\vrule height0.5\ht0\hss}\box0}}
{\setbox0=\hbox{$\textstyle        \rm S$}\hbox{\raise0.5\ht0\hbox
to0pt{\kern0.35\wd0\vrule height0.45\ht0\hss}\hbox
to0pt{\kern0.55\wd0\vrule height0.5\ht0\hss}\box0}}
{\setbox0=\hbox{$\scriptstyle      \rm S$}\hbox{\raise0.5\ht0\hbox
to0pt{\kern0.35\wd0\vrule height0.45\ht0\hss}\raise0.05\ht0\hbox
to0pt{\kern0.5\wd0\vrule height0.45\ht0\hss}\box0}}
{\setbox0=\hbox{$\scriptscriptstyle\rm S$}\hbox{\raise0.5\ht0\hbox
to0pt{\kern0.4\wd0\vrule height0.45\ht0\hss}\raise0.05\ht0\hbox
to0pt{\kern0.55\wd0\vrule height0.45\ht0\hss}\box0}}}}
\def\bbbz{{\mathchoice {\hbox{$\sans\textstyle Z\kern-0.4em Z$}}
{\hbox{$\sans\textstyle Z\kern-0.4em Z$}}
{\hbox{$\sans\scriptstyle Z\kern-0.3em Z$}}
{\hbox{$\sans\scriptscriptstyle Z\kern-0.2em Z$}}}}
%-----------------------------------------------------------------------
% petit-fonts
\skewchar\ninei='177 \skewchar\sixi='177
\skewchar\ninesy='60 \skewchar\sixsy='60
\hyphenchar\ninett=-1
\def\newline{\hfil\break}%
%-----------------------------------------------------------------------
\catcode`@=11
\def\folio{\ifnum\pageno<\z@
\uppercase\expandafter{\romannumeral-\pageno}%
\else\number\pageno \fi}
\catcode`@=12 % at signs are no longer letters
%-------------------------------------------------------
% Definition der versal griechischen Buchstaben
%=======================================================================
  \mathchardef\Gamma="0100
  \mathchardef\Delta="0101
  \mathchardef\Theta="0102
  \mathchardef\Lambda="0103
  \mathchardef\Xi="0104
  \mathchardef\Pi="0105
  \mathchardef\Sigma="0106
  \mathchardef\Upsilon="0107
  \mathchardef\Phi="0108
  \mathchardef\Psi="0109
  \mathchardef\Omega="010A
%-----------------------------------------------------------------------
\def\squareforqed{\hbox{\rlap{$\sqcap$}$\sqcup$}}
\def\qed{\ifmmode\squareforqed\else{\unskip\nobreak\hfil
\penalty50\hskip1em\null\nobreak\hfil\squareforqed
\parfillskip=0pt\finalhyphendemerits=0\endgraf}\fi}
%-----------------------------------------------------------------------
\newfam\sansfam
\textfont\sansfam=\tensans\scriptfont\sansfam=\sevensans
\scriptscriptfont\sansfam=\fivesans
\def\sans{\fam\sansfam\tensans}
%-----------------------------------------------------------------------
\def\stackfigbox{\if
Y\FIG\global\setbox\figbox=\vbox{\unvbox\figbox\box1}%
\else\global\setbox\figbox=\vbox{\box1}\global\let\FIG=Y\fi}
\def\placefigure{\dimen0=\ht1\advance\dimen0by\dp1
\advance\dimen0by5\baselineskip
\advance\dimen0by0.4true cm
\ifdim\dimen0>\vsize\pageinsert\box1\vfill\endinsert
\else%keine seitenhohe Abbildung
\if Y\FIG\stackfigbox\else
\dimen0=\pagetotal\ifdim\dimen0<\pagegoal%akt. Seite ist noch nicht voll
\advance\dimen0by\ht1\advance\dimen0by\dp1\advance\dimen0by1.7true cm
\ifdim\dimen0>\pagegoal\stackfigbox
\else\box1\vskip7true mm\fi
\else\box1\vskip7true mm\fi\fi\fi\let\firstleg=Y}
%
% Abbildungen
\def\begfig#1cm#2\endfig{\par
\setbox1=\vbox{\dimen0=#1true cm\advance\dimen0
by1true cm\kern\dimen0\vskip-.8333\baselineskip#2}\placefigure}
\def\begdoublefig#1cm #2 #3 \enddoublefig{\begfig#1cm%
\line{\vtop{\hsize=0.46\hsize#2}\hfill
\vtop{\hsize=0.46\hsize#3}}\endfig}
%-------------------------------------------------------------------
\let\firstleg=Y
% Abbildungslegenden
% Falls Text kleiner als eine volle Zeile, zentriert.
\def\figure#1#2{\if Y\firstleg\vskip1true cm\else\vskip1.7true mm\fi
\let\firstleg=N\setbox0=\vbox{\noindent\petit{\bf
Fig.\ts#1\unskip.\ }\ignorespaces #2\smallskip
\count255=0\global\advance\count255by\prevgraf}%
\ifnum\count255>1\box0\else
\centerline{\petit{\bf Fig.\ts#1\unskip.\
}\ignorespaces#2}\smallskip\fi}
%-----------------------------------------------------------------
% Tabellenkoepfe

%-------------------------------------------------------------------
\def\begtab#1cm#2\endtab{\par
   \ifvoid\topins\midinsert\medskip\vbox{#2\kern#1true cm}\endinsert
   \else\topinsert\vbox{#2\kern#1true cm}\endinsert\fi}
%-------------------------------------------------------------------
\def\begpet{\vskip6pt\bgroup\petit}
\def\endpet{\vskip6pt\egroup}
%-------------------------------------------------------------------
% Referenzen
\newdimen\refindent
\newlinechar=`\^
\def\begref#1#2{\titlea{}{#1}%
\bgroup\petit
\setbox0=\hbox{#2\enspace}\refindent=\wd0\relax
\if!#2!\else
\ifdim\refindent>0.5em\else
\message{^Something may be wrong with your references;}%
\message{probably you missed the second argument of \string\begref.}%
\fi\fi}
\def\ref{\goodbreak
\hangindent\oldparindent\hangafter=1
\noindent\ignorespaces}
\def\refno#1{\goodbreak
\setbox0=\hbox{#1\enspace}\ifdim\refindent<\wd0\relax
\message{^Your reference `#1' is wider than you pretended in using
\string\begref.}\fi
\hangindent\refindent\hangafter=1
\noindent\kern\refindent\llap{#1\enspace}\ignorespaces}
\def\refmark#1{\goodbreak
\setbox0=\hbox{#1\enspace}\ifdim\refindent<\wd0\relax
\message{^Your reference `#1' is wider than you pretended in using
\string\begref.}\fi
\hangindent\refindent\hangafter=1
\noindent\hbox to\refindent{#1\hss}\ignorespaces}
\def\endref{\goodbreak\endpet}% Ende der Referenzen
%-------------------------------------------------------------------
\def\vec#1{{\textfont1=\tenbf\scriptfont1=\sevenbf
\textfont0=\tenbf\scriptfont0=\sevenbf
\mathchoice{\hbox{$\displaystyle#1$}}{\hbox{$\textstyle#1$}}
{\hbox{$\scriptstyle#1$}}{\hbox{$\scriptscriptstyle#1$}}}}
%---------------------------------------------------------------------
\def\petit{\def\rm{\fam0\ninerm}%
\textfont0=\ninerm \scriptfont0=\sixrm \scriptscriptfont0=\fiverm
 \textfont1=\ninei \scriptfont1=\sixi \scriptscriptfont1=\fivei
 \textfont2=\ninesy \scriptfont2=\sixsy \scriptscriptfont2=\fivesy
 \def\it{\fam\itfam\nineit}%
 \textfont\itfam=\nineit
 \def\sl{\fam\slfam\ninesl}%
 \textfont\slfam=\ninesl
 \def\bf{\fam\bffam\ninebf}%
 \textfont\bffam=\ninebf \scriptfont\bffam=\sixbf
 \scriptscriptfont\bffam=\fivebf
 \def\sans{\fam\sansfam\ninesans}%
 \textfont\sansfam=\ninesans \scriptfont\sansfam=\sixsans
 \scriptscriptfont\sansfam=\fivesans
 \def\tt{\fam\ttfam\ninett}%
 \textfont\ttfam=\ninett
 \normalbaselineskip=11pt
 \setbox\strutbox=\hbox{\vrule height7pt depth2pt width0pt}%
 \normalbaselines\rm
\def\vec##1{{\textfont1=\tbms\scriptfont1=\tbmss
\textfont0=\ninebf\scriptfont0=\sixbf
\mathchoice{\hbox{$\displaystyle##1$}}{\hbox{$\textstyle##1$}}
{\hbox{$\scriptstyle##1$}}{\hbox{$\scriptscriptstyle##1$}}}}}
%-------------------------------------------------------------------
\nopagenumbers
%
% Der Schalter \header gibt an, ob ein "running head" gedruckt werden
% soll; wenn er nicht auf "N" steht kommt ein solcher.
\let\header=Y
\let\FIG=N
\newbox\figbox
\output={\if N\header\headline={\hfil}\fi\plainoutput
\global\let\header=Y\if Y\FIG\topinsert\unvbox\figbox\endinsert
\global\let\FIG=N\fi}
%------------------------------------------------------
\let\lasttitle=N
%------------------------------------------------------
%iukt-changes: centerpar from lecproc.cmm included,
% used in author, address, contribution
%------------------------------------------------------
\def\centerpar#1{{\parfillskip=0pt
\rightskip=0pt plus 1fil
\leftskip=0pt plus 1fil
\advance\leftskip by\oldparindent
\advance\rightskip by\oldparindent
\def\newline{\break}%
\noindent\ignorespaces#1\par}}
%---------------------------------------------------------------
\catcode`\@=\active
\def\author#1{\bgroup
\baselineskip=13.2pt
\lineskip=0pt
\pretolerance=10000
\markfont
\centerpar{#1}\bigskip\egroup
{\def@##1{}%
\setbox0=\hbox{\petit\kern2.5true cc\ignorespaces#1\unskip}%
\ifdim\wd0>\hsize
\message{The names of the authors exceed the headline, please use a }%
\message{short form with AUTHORRUNNING}\gdef\leftheadline{%
\rlap{\folio}\hfil AUTHORS suppressed due to excessive length}%
\else
\xdef\leftheadline{\rlap{\noexpand\folio}\hfil
\ignorespaces#1\unskip}%
\fi
}\let\INS=E}
\def\address#1{\bgroup\petit
\centerpar{#1}\bigskip\egroup
\catcode`\@=12
\vskip2cm\noindent\ignorespaces}
%---------------------------------------------------------------------
\let\INS=N%
% Aktionen, die bei Antreffen des @-Zeichens zu machen sind;
% drei Faelle a) @ bei AUTHOR, b) 1.@ bei ADDRESS, c) alle weiteren @'s
\def@#1{\if N\INS\unskip$\,^{#1}$\else\global\footcount=#1\relax
\if E\INS\hangindent0.5\parindent\noindent\hbox
to0.5\parindent{$^{#1}$\hfil}\let\INS=Y\ignorespaces
\else\par\hangindent0.5\parindent\noindent\hbox
to0.5\parindent{$^{#1}$\hfil}\ignorespaces\fi\fi}%
\catcode`\@=12
%-------------------------------------------------------------------
% "running head"
\headline={\petit\def\newline{ }\def\fonote#1{}\ifodd\pageno
\rightheadline\else\leftheadline\fi}
\def\rightheadline{Missing CONTRIBUTION
title\hfil\llap{\folio}}
\def\leftheadline{\rlap{\folio}\hfil Missing name(s)
of the author(s)}
\nopagenumbers
\let\header=Y
%------------------------------------------------------

%------------------------------------------------------
\let\lasttitle=N
 \def\contribution#1{\vfill\eject
 \let\header=N\bgroup
 \textfont0=\tafontt \scriptfont0=\tafonts \scriptscriptfont0=\tafontss
 \textfont1=\tamt \scriptfont1=\tams \scriptscriptfont1=\tams
 \textfont2=\tast \scriptfont2=\tass \scriptscriptfont2=\tasss
 \par\baselineskip=16pt
     \lineskip=16pt
     \tafontt
     \raggedright
     \pretolerance=10000
     \noindent
     \centerpar{\ignorespaces#1}%
     \vskip17pt\egroup
     \nobreak
     \parindent=0pt
     \everypar={\global\parindent=1.5em
     \global\let\lasttitle=N\global\everypar={}}%
     \global\let\lasttitle=A%
     \setbox0=\hbox{\petit\def\newline{ }\def\fonote##1{}\kern2.5true
     cc\ignorespaces#1}\ifdim\wd0>\hsize
     \message{Your CONTRIBUTIONtitle exceeds the headline,
please use a short form
with CONTRIBUTIONRUNNING}\gdef\rightheadline{CONTRIBUTION title
suppressed due to excessive length\hfil\llap{\folio}}%
\else
\gdef\rightheadline{\ignorespaces#1\unskip\hfil\llap{\folio}}\fi
\catcode`\@=\active
     \ignorespaces}
%------------------------------------------------------
% Beginn Ueberschrift 1. Ordnung
\def\titlea#1#2{\if N\lasttitle\else\vskip-28pt
     \fi
     \vskip18pt plus 4pt minus4pt
     \bgroup
\textfont0=\tbfontt \scriptfont0=\tbfonts \scriptscriptfont0=\tbfontss
\textfont1=\tbmt \scriptfont1=\tbms \scriptscriptfont1=\tbmss
\textfont2=\tbst \scriptfont2=\tbss \scriptscriptfont2=\tbsss
\textfont3=\tasys \scriptfont3=\tenex \scriptscriptfont3=\tenex
     \baselineskip=16pt
     \lineskip=0pt
     \pretolerance=10000
     \noindent
     \tbfontt
     \rightskip 0pt plus 6em
     \setbox0=\vbox{\vskip23pt\def\fonote##1{}%
     \noindent
     \if!#1!\ignorespaces#2
     \else\ignorespaces#1\unskip\enspace\ignorespaces#2\fi
     \vskip18pt}%
     \dimen0=\pagetotal\advance\dimen0 by-\pageshrink
     \ifdim\dimen0<\pagegoal
     \dimen0=\ht0\advance\dimen0 by\dp0\advance\dimen0 by
     3\normalbaselineskip
     \advance\dimen0 by\pagetotal
     \ifdim\dimen0>\pagegoal\eject\fi\fi
     \noindent
     \if!#1!\ignorespaces#2
     \else\ignorespaces#1\unskip\enspace\ignorespaces#2\fi
     \vskip12pt plus4pt minus4pt\egroup
     \nobreak
     \parindent=0pt
     \everypar={\global\parindent=\oldparindent
     \global\let\lasttitle=N\global\everypar={}}%
     \global\let\lasttitle=A%
     \ignorespaces}
%------------------------------------------------------
 % Beginn Ueberschrift 2. Ordnung
 \def\titleb#1#2{\if N\lasttitle\else\vskip-22pt
     \fi
     \vskip18pt plus 4pt minus4pt
     \bgroup
\textfont0=\tenbf \scriptfont0=\sevenbf \scriptscriptfont0=\fivebf
\textfont1=\tams \scriptfont1=\tamss \scriptscriptfont1=\tbmss
     \lineskip=0pt
     \pretolerance=10000
     \noindent
     \bf
     \rightskip 0pt plus 6em
     \setbox0=\vbox{\vskip23pt\def\fonote##1{}%
     \noindent
     \if!#1!\ignorespaces#2
     \else\ignorespaces#1\unskip\enspace\ignorespaces#2\fi
     \vskip10pt}%
     \dimen0=\pagetotal\advance\dimen0 by-\pageshrink
     \ifdim\dimen0<\pagegoal
     \dimen0=\ht0\advance\dimen0 by\dp0\advance\dimen0 by
     3\normalbaselineskip
     \advance\dimen0 by\pagetotal
     \ifdim\dimen0>\pagegoal\eject\fi\fi
     \noindent
     \if!#1!\ignorespaces#2
     \else\ignorespaces#1\unskip\enspace\ignorespaces#2\fi
     \vskip8pt plus4pt minus4pt\egroup
     \nobreak
     \parindent=0pt
     \everypar={\global\parindent=\oldparindent
     \global\let\lasttitle=N\global\everypar={}}%
     \global\let\lasttitle=B%
     \ignorespaces}
%------------------------------------------------------
 % Beginn Ueberschrift 3. Ordnung
 \def\titlec#1{\if N\lasttitle\else\vskip-\baselineskip
     \fi
     \vskip18pt plus 4pt minus4pt
     \bgroup
\textfont0=\tenbf \scriptfont0=\sevenbf \scriptscriptfont0=\fivebf
\textfont1=\tams \scriptfont1=\tamss \scriptscriptfont1=\tbmss
     \bf
     \noindent
     \ignorespaces#1\unskip\ \egroup
     \ignorespaces}
%-------------------------------------------------------------------
 % Beginn Ueberschrift 4. Ordnung
 \def\titled#1{\if N\lasttitle\else\vskip-\baselineskip
     \fi
     \vskip12pt plus 4pt minus 4pt
     \bgroup
     \it
     \noindent
     \ignorespaces#1\unskip\ \egroup
     \ignorespaces}
%-------------------------------------------------------------------
\let\ts=\thinspace
\def\footnoterule{\kern-3pt\hrule width 2true cm\kern2.6pt}
% Fussnoten-macros
\newcount\footcount \footcount=0
\def\advftncnt{\advance\footcount by1\global\footcount=\footcount}
% Automatisch numerierte Fussnote, Fussnotentex in petit
\def\fonote#1{\advftncnt$^{\the\footcount}$\begingroup\petit
\parfillskip=0pt plus 1fil
\def\textindent##1{\hangindent0.5\oldparindent\noindent\hbox
to0.5\oldparindent{##1\hss}\ignorespaces}%
\vfootnote{$^{\the\footcount}$}{#1\vskip-9.69pt}\endgroup}
%-------------------------------------------------------------------
\def\item#1{\par\noindent
\hangindent6.5 mm\hangafter=0
\llap{#1\enspace}\ignorespaces}
%-------------------------------------------------------------------

%-------------------------------------------------------------------
\def\newenvironment#1#2#3#4{\long\def#1##1##2{\removelastskip
\vskip\baselineskip\noindent{#3#2\if!##1!.\else\unskip\ \ignorespaces
##1\unskip\fi\ }{#4\ignorespaces##2}\vskip\baselineskip}}
% Lemma, Proposition, Theorem, Corollary
\newenvironment\lemma{Lemma}{\bf}{\it}
\newenvironment\proposition{Proposition}{\bf}{\it}
\newenvironment\theorem{Theorem}{\bf}{\it}
\newenvironment\corollary{Corollary}{\bf}{\it}
%---------------------------------------------------------------------
% Example, Exercise, Problem, Solution, Definition
\newenvironment\example{Example}{\it}{\rm}
\newenvironment\exercise{Exercise}{\bf}{\rm}
\newenvironment\problem{Problem}{\bf}{\rm}
\newenvironment\solution{Solution}{\bf}{\rm}
\newenvironment\definition{Definition}{\bf}{\rm}
%---------------------------------------------------------------------
%Note, Question
\newenvironment\note{Note}{\it}{\rm}
\newenvironment\question{Question}{\it}{\rm}
%---------------------------------------------------------------------
%Proof, Remark
\long\def\remark#1{\removelastskip\vskip\baselineskip\noindent{\it
Remark.\ }\ignorespaces}
\long\def\proof#1{\removelastskip\vskip\baselineskip\noindent{\it
Proof\if!#1!\else\ \ignorespaces#1\fi.\ }\ignorespaces}
%------------------------------------------------------------------
\def\typeset{\petit\noindent This article was processed by the author
using the \TeX\ macro package from Springer-Verlag.\par}
\outer\def\byebye{\bigskip\bigskip\typeset
\footcount=1\ifx\speciali\undefined\else
\loop\smallskip\noindent special character No\number\footcount:
\csname special\romannumeral\footcount\endcsname
\advance\footcount by 1\global\footcount=\footcount
\ifnum\footcount<11\repeat\fi
\vfill\supereject\end}

% This is MATHMULT.DEM the demonstration file of
% the plain TeX macro package from Springer-Verlag
% for multi author books in mathematics, version of March 1991
\def\12{{1\ov 2}}

\def\ov{\over}

%\input mathmult.cmm
% !!!!!!!! here starts the real stuff !!!!!!!!!!!!!!!!!!!!!!!!!!!
\def\Tr{\,{\rm Tr}\,}

\contribution{Electroweak Reactions in the Non-Perturbative Regime
of QCD}
%\author{Ulf-G. Mei{\ss}ner@1}
%\address{@1University of Berne, Institute for Theoretical Physics,
\author{Ulf-G. Mei{\ss}ner}
\footnote{}{\vskip -0.6truecm Lectures delivered at the XXXII. Internationale
Universit\"atswochen f\"ur Kern- und Teilchenphysik, Schladming, Styria,
Austria, February 24 - March 6, 1993.}
\address{University of Berne, Institute for Theoretical Physics,
CH-3012 Berne, Switzerland}
% List of Contents - use for preprint or separate for the book ********
\vskip -2truecm
\titlea{0}{Contents}
\item{1} Introduction \hfill 1
\medskip
\item{2} Chiral Perturbation Theory with Nucleons \hfill 2
\smallskip
2.1 \quad Chiral Symmetry in QCD \hfill 2

2.2 \quad Chiral Perturbation Theory (Mesons)  \hfill 3

2.3 \quad Inclusion of Matter Fields  \hfill 5

2.4 \quad Threshold Pion Photo- and Electroproduction  \hfill 7

2.5 \quad Nucleon Compton Scattering  \hfill 9

\medskip
\item{3} The Quark Structure of the Nucleon \hfill 12
\smallskip
3.1 \quad Currents, Sizes and Form Factors of the Nucleon  \hfill 12

3.2 \quad A Toy Model: Quark Distributions from Form Factors \hfill 14

3.3 \quad Electroweak Currents \hfill 15

3.4 \quad Parity--Violating Electron Scattering  \hfill 18

3.5 \quad Neutrino and Antineutrino Scattering off Nucleons and Nuclei \hfill
19

3.6 \quad Down and Dirty: QED, QCD and Heavy Quark Corrections \hfill 21

\medskip
\item{4} Summary and Outlook  \hfill 25
\titlea{1}{Introduction}
These lectures are concerned with the structure of hadrons at low
energies, where the strong coupling constant is large. Most of the
hadronic world discussed here will be made up of the light u, d and s
quarks since these are the constituents of the low-lying hadrons. The
best way to gain information about the strongly interacting particles
is thus the use of well-understood probes, such as the photon or the
massive weak gauge bosons. At very low energies, the dynamics of the
strong interactions is governed by constraints from chiral symmetry.
This leads to the use of effective field theory methods which in the
present context is called baryon chiral perturbation theory. In the
first part of these lectures, I will briefly outline the basic framework
of this effective field theory and use photo-nucleon processes to
discuss the strengths and limitations of it. The basic degrees of
freedom are the pseudoscalar Goldstone bosons chirally coupled to the
matter fields like e.g. the nucleons. The very low-energy face of the
low-lying baryons is therefore of hadronic nature, essentially
point-like Dirac particles surrounded by a cloud of Goldstone bosons.
The information about the underlying quark structure is only indirect
since one translates e.g. quark masses into pseudoscalar meson masses.
Nevertheless, certain aspects of the quark structure of the nucleon
(or other baryons) can be extracted by making clever use of the
well-known couplings of the quarks to the electroweak gauge bosons.
For that, it is important to understand the notion of hadron form
factors, electromagnetic or weak ones. These encode certain aspects
of the quark structure of the nucleon. In addition, parity-violating
electron scattering off nucleons or nuclei or neutrino-nucleon
scattering allows one to measure the weak form factors related to the
neutral currents which complement the information gained from the
electromagnetic and weak charged current form factors. This will be
the theme of the second part of these lectures. However, since one is
dealing with small effects like e.g. left-right asymmetries in
polarized electron scattering off protons at low energies, one also
has to worry about the effects of the heavy quarks c, b and t and about
radiative corrections, which are nominally suppressed by powers of
($\alpha / 4 \pi)$ but sometimes enhanced by large logarithms or by
some judiciously large factors. I will give a short state of the art
summary on these topics at the end of these lectures.

Naturally, I can only cover a small fraction of the many interesting
phenomena related to low energy hadron physics. I have chosen to
mostly talk about the nucleon since after all it makes up large
chunks of the stable matter surrounding us and also is a good
intermediary between the nuclear and the high energy physicists
present at this school. Most of the methods presented here can easily
be applied to other problems, and as it will become obvious at many
places, we still have a long way to go to understand all the intriguing
features of the nucleon in a systematic and controlled fashion. Whenever
possible, I will avoid to talk about models, with the exception of some
circumstances where they offer some qualitative insight into certain
aspects of hadron structure.

\titlea{2}{Chiral Perturbation Theory with Nucleons}
\titleb{2.1}{Chiral Symmetry in QCD}
Consider the QCD Lagrangian for the three light flavors u, d and s,
$${\cal L}_{QCD} = \bar q (i \gamma_\mu D^\mu - {\cal M}) q
+ {\cal L}_{{\rm gluon}} \eqno(2.1) $$
with $q =$ (u, d, s) and I have suppressed all color indices. The
quark fields can be chosen such that the quark mass matrix is
diagonal,
$$ {\cal M} = \left(\matrix{  m_u&& \cr &m_d&  \cr &&m_s \cr}\right)
\eqno(2.2) $$
The entries in (2.2) are the current quark masses. At the typical
hadronic scale of $\Lambda \simeq 1$ GeV, these can be considered
small. This holds certainly for the u and d quarks, the s quark is
somewhat heavier ($\simeq 175$ MeV) which makes it more difficult to
deal with. To a good first approximation, it is, however, legitimate
to set the current quark masses to zero. In that case, the QCD
Lagrangian exhibits a flavor $SU(3)_L \times SU(3)_R \times U(1)_{L+R}
\times U(1)_{L-R}$ symmetry. The vectorial $U(1)$ is related to the
baryon number and the axial $U(1)$ is afflicted by the anomaly which
gives the $\eta'$ a mass of 1 GeV. The chiral group has therefore
sixteen independent generators. This leads to sixteen conserved
left- and right-handed currents,
$$ J_{\mu,A}^a = \bar q_A \gamma_\mu {\lambda^a \over 2} q_A
\quad (a=1, \ldots , 8 \, ; \,  A= L,R)      \eqno(2.3) $$
or, equivalently, eight
conserved vector and eight conserved axial-vector
charges. These form an SU(3) algebra. However, this symmetry of the
Lagrangian is not present in the ground state or the spectrum of the
strongly interacting particles. Something very similar to the
ferromagnet below the Curie temperature happens. The magnet Hamiltonian
is rotationally invariant but below $T_c$ spontaneous magnetization
sets in and all spins align in one direction breaking the symmetry
of the Hamiltonian. Similarly in QCD only the vector charges are
conserved allowing us to order the hadrons in multiplets of certain
isospin and hypercharge quantum numbers. Nevertheless, the information
about the axial charges is not lost. Goldstone's theorem [1] tells
us that for any broken generator there exists a massless boson, in
this case of pseudoscalar nature. This means that we should expect
eight massless pseudoscalar mesons in the particle data listings.
This is not the case. The fact of the matter is that the Goldstone
bosons pick up a small mass related to the small current quark masses,
like e.g.
$$ M_\pi^2 = (m_u + m_d) B + O({\cal M}^2)     \eqno(2.4) $$
where $ B=-<0 | \bar q q | 0> / F_\pi^2$ is the order parameter of the
spontaneous symmetry breaking and $F_\pi$ measures the strength of
the non-vanishing transition amplitude
$$<0 | A_\mu^i | \pi^j> = i \delta^{ij} p_\mu F_\pi  \, \, .
\eqno(2.5)$$
In what follows, I will assume that $B \sim 1$ GeV, i.e. that the
quark mass expansion of the pseudoscalar meson masses is indeed
governed by the leading term linear in the quark masses. It is
interesting to observe that (2.4) gives us a recipe how to translate
quark into meson masses. Evidently, at low energies the almost
massless Goldstone bosons are the dominant degrees of freedom. This
has been the theme of Heiri Leutwyler's lectures [2] two years ago
and I refer the reader for a more detailed discussion of these
underlying ideas. Also a recent review [3] is available which can be
used as a first reading.
\titleb{2.2}{Chiral Perturbation Theory (Mesons)}
The purest reaction of low-energy QCD is the elastic scattering of pions from
pions in the threshold region. I will use this as an example to discuss the
principles underlying the so-called chiral perturbation theory (CHPT). It is
convenient to collect the Goldstone bosons in a matrix-valued field $U(x) =
\exp{ (i \phi (x) / F_\pi) }$,
in the standard notation. To lowest order in the
momenta and quark masses, the interactions of the Goldstone bosons e.g. can be
described by the gauged non-linear $\sigma$-model,
$${\cal L}^{(2)} = { F^2 \over 4} {\rm Tr} [\nabla_\mu U^\dagger \nabla^\mu U
] +  {F^2 B\over 4} {\rm Tr} [{\cal M}( U + U^\dagger )] \, \, , \eqno(2.6)$$
with $\nabla_\mu$ a covariant derivative including the coupling to the
external electroweak fields. The second term in (2.6) reflects the explicit
symmetry breaking from the quark masses. Both terms are of chiral dimension
two since they contain either two derivatives or one quark mass (which is
equivalent to two derivatives, see (2.4)). To lowest order, the strong
interactions are therefore given in terms of two parameters. These are $F$,
the pion decay constant in the chiral limit, and $B$ related to the quark
condensate (remember that I assume $B$ to be substantially larger than $F$).
Calculating tree diagrams from this effective Lagrangian leads to the
celebrated current algebra results. In particular, one recovers Weinberg's
prediction for the $\pi \pi$ scattering amplitude, $A(s,t,u) = (s - M_\pi^2 )
/ F_\pi^2$ [4]. Clearly, this smells like an expansion in energy since close
to threshold, the c.m. energy $\sqrt{s}$ becomes very small and, consequently,
the interaction between  the Goldstone bosons weak. This is a particular
imprint of the spontaneously broken chiral symmetry which allows for a
"perturbative" expansion in the non-perturbative regime of QCD. Obviously, the
effective Lagrangian (2.6) can not be the whole story. First, one would like
to know how big the corrections from higher order terms (with more derivatives
and/or quark mass insertions) are. Second, and more important, tree diagrams
are always real. Twenty years ago at this school, Lehmann [5] demonstrated
that unitarity forces one to consider loop diagrams. Based on the counting
rules developed by Weinberg [6], which state that $n$-loop diagrams are
suppressed by powers of $q^{2n}$ (with $q$ a genuine small momentum), Gasser
and Leutwyler [7] have systematized the whole approach for QCD. Let us
consider the corrections which appear at next-to-leading order. First, there
are the one-loop diagrams mandated by unitarity. Second, there are higher
order contact terms with more derivatives or quark mass insertions. These are
accompanied by a priori unknown coupling constants, the so-called low-energy
constants. Their values can be fixed from phenomenology [7] or understood to a
high degree of accuracy from the QCD version of the vector-meson dominance
principle [8]. It is at this next-to-leading order where the various
equivalent lowest order approaches differ. For example, the linear
$\sigma$-model is a perfect candidate to be used instead of (2.6), but at
next-to-leading order it simply gives low-energy constants at variance with
the data [7]. Let me return to the example of $\pi \pi$ scattering. The
amplitude expanded to one loop accuracy takes the form
$$T_{\pi \pi} (s,t,u) = A(s,t,u) + B(s,t,u) + C(s,t,u) \, \, , \eqno(2.7)$$
where $B(s,t,u)$ includes the loop (unitarity) effects and $C(s,t,u)$ the
contribution from the higher dimensional operators. That this gives indeed a
better description of the $\pi \pi$ phase shifts than the tree level
prediction $A(s,t,u)$ can be seen in ref.[9]. In what follows, I will always
confine myself to the one loop approximation. For a discussion about its range
of applicability, see e.g. ref.[10].
\titleb{2.3}{Inclusion of Matter Fields}
In this section I will be concerned with adding the low-lying spin-1/2 baryons
($N$, $\Sigma$, $\Lambda$, $\Xi$) to the effective field theory. The inclusion
of such matter fields is less straightforward since these particles are not
related to the symmetry violation. However, their interactions with the
Goldstone bosons is dictated by chiral symmetry. Let us denote by $B$ the
conventional $3 \times 3$ matrix representing the spin-1/2 octet fields,
$$ B =  \left(
\matrix  { {1\over \sqrt 2} \Sigma^0 + {1 \over \sqrt 6} \Lambda
&\Sigma^+ &  p \cr
\Sigma^-
    & -{1\over \sqrt 2} \Sigma^0 + {1 \over \sqrt 6} \Lambda & n \cr
\Xi^-
        &       \Xi^0 &- {2 \over \sqrt 6} \Lambda \cr} \right)
\eqno(2.8) $$
It is most convenient to choose a non-linear realization  of the chiral
symmetry [11] so that $B$ transforms as
$$ B \to K B K^\dagger
\eqno(2.9)$$
where $K$ is a complicated function that does not only depend on the group
elements $g_{L,R}$ of the $SU(3)_{L,R}$ but also on the Goldstone boson fields
collected in $U(x)$, i.e. $K(x) = K( g_L , g_R , U(x) )$ defines a local
transformation. Expanding $K$ in powers of the Goldstone boson fields, one
realizes that a chiral transformation is linked to absorption or emission of
pions, kaons and etas (which was the theme in the days of "current algebra"
techniques). These topics are discussed in more detail in [3]. The lowest
order meson-baryon Lagrangian is readily constructed. Let us restrict the
discussion to processes with one incoming and one outgoing baryon, such as
$\pi N$ scattering, threshold pion photoprodcution or baryon Compton
scattering (otherwise, we would have to add contact $n$-fermion terms with $n
\ge 4$). In that case, it takes the form
$${\cal L}_{\rm MB}^{(1)} = \Tr \bigl\lbrace i \bar{B} \gamma^\mu D_\mu B
-     m_0  \bar{B} B + {1 \over 2} D \bar{B} \gamma^\mu \gamma_5
\lbrace
u_\mu , B \rbrace
+ {1 \over 2} F \bar{B} \gamma^\mu \gamma_5 [ u_\mu , B ] \bigr\rbrace
\eqno(2.10)$$
with $m_0$ the average octet mass (in the chiral limit) and $D_\mu$ the
usual covariant derivative. There are two axial-vector type couplings
multiplied by the conventional $F$ and $D$ constants. In the case of two
flavors, there is only one such term proportional to $g_A = F + D$. Notice
that the lowest order effective Lagrangian contains one derivative and
therefore is of dimension one as indicated by the superscript '(1)'. In
contrast to the meson sector (2.6), odd powers of the small momentum $q$ are
allowed (thus, to leading order, no quark mass insertion appears since ${\cal
M} \sim q^2$). It is instructive to expand (2.10) in powers of the Goldstone
and external fields. From the vectorial term, one gets the minimal
photon-baryon coupling, the two-Goldstone seagull (Weinberg term) and many
others. Expansion of the axial-vectors leads to the pseudovector meson-baryon
coupling, the celebrated Kroll-Rudermann term and much more. However, as
already stated, it is not sufficient to calculate tree diagrams from the
lowest order effective theory. In the presence of baryons, the loop expansion
is more complicated than discussed in the previous section. First, since odd
powers in $q$ are allowed, a one-loop calculation of order $q^3$ involves
contact terms of dimension two and three, i.e. combinations of zero or one
quark mass insertions with zero to three derivatives. These terms are
collected in ${\cal L}_{\rm MB}^{(2,3)}$ and a complete list of them can be
found in Krause's paper [12]. Second, the finiteness of the baryon mass in the
chiral limit and that its value is comparable to the chiral symmetry breaking
scale $\Lambda \sim M_\rho$ complicates the low energy structure. This has
been discussed in detail by Gasser et al. [13]. Let me just give one
illustrative example. The one loop contribution  to the nucleon mass not only
gives the celebrated non-analytic contribution proportional to $M_\pi^3 \sim
{\cal M}^{3/2}$ but also an infinite shift of $m_0$ which has to be
compensated by a counterterm of dimension zero. It is a general feature that
loops produce analytic contributions at orders below what one would naively
expect (e.g. below $q^3$ from one loop diagrams). Therefore, in a CHPT
calculation involving baryons one has to worry more about higher order
contributions than it is the case in the meson sector. There is one way of
curing this problem, namely to go into the extreme non-relativistic limit [14]
and consider the baryons as very heavy (static) sources. Then, by a clever
definition of velocity-dependent fields, one can eliminate the baryon mass
term from the lowest order effective Lagrangian and expand all interaction
vertices and baryon propagators in increasing powers of $1 / m_0$. This is
similar to a Foldy-Wouthuysen transformation you all know from QED. In this
limit one recovers a consistent derivative expansion. For example, the one
loop contribution of the Goldstone bosons to the baryon self-energy is nothing
but the non-analytic $M_\phi^3$ terms
together with three contact terms from
${\cal L}_{\rm MB}^{(2)}$. However, one has to be somewhat careful still. The
essence of the heavy mass formalism is that one works with old-fashioned
time-ordered perturbation theory. So one has to watch out for the appearance
of possible small energy denominators (infrared singularities). This problem
has been addressed by Weinberg [15] in his discussion about the nature of the
nuclear forces. The dangerous diagrams are the ones  were cutting one pion
line (this only concerns pions which are not in the asymptotic in- or
out-states)
separates the diagram into two disconnected pieces (one therefore speaks
of reducible diagrams). These diagrams should be inserted in a Schr\"odinger
equation or a relativistic generalization thereof with the irreducible ones
entering as a potential. So the full CHPT machinery is applied to the
irreducible diagrams. This should be kept in mind. For the purposes I am
discussing, we do not need to worry about these complications. Being aware of
them, it is then straightforward to apply baryon CHPT to many nuclear and
particle physics problems [3,16,17]. I will illustrate this on two particular
examples in the next sections. Before doing that, however, I would like to
stress that all these calculations are only in their infancy. It is believed
that for a good quantitative description one has to perform  systematic
calculations to order $q^4$, i.e. beyond next-to-leading order. There are many
indications of this coming from the baryon masses, hyperon non-leptonic decays
and so on. At present, only some rather unsystematic attempts have been done
which incorporate some $q^4$ effects. In particular, it has been argued that
one should include the low-lying spin-3/2 decuplet in the effective theory
from the start [16,18]. A critical discussion of this approach can be
found in ref.[19]. Clearly, more work is needed to clarify these issues. Let
me now proceed and discuss two applications of the framework outlined here.
These are calculations in flavor $SU(2)$ were the loop corrections are
considerably smaller. As I will show, they shed some light on the
non-perturbative structure of the nucleon.
\titleb{2.4}{Threshold Pion Photo- and Electroproduction}
Let us consider the reaction $\gamma (k) + N_1 (p_1) \to \pi^a (q) + N_2 (p_2)$
with $N_{1,2}$ denoting protons and/or neutrons and '$a$' refers to the charge
of the produced pion. In the case of real photons ($k^2 = 0$) one talks of
photoproduction whereas for virtual photons (radiated off an electron beam)
the process is called electroproduction. Of particular interest is the
threshold region where the photon has just enough energy to produce the pion at
rest or with a very small three-momentum. In this kinematical regime, it is
advantegous to perform a multipole decomposition since at threshold only the
$S$-waves survive. These multipoles are labelled $E \, (M)_{l \pm}$, with $E \,
(M)$ for electric (magnetic), $l = 0,1,2, \ldots$ the pion orbital angular
momentum and
the $\pm$ refers to the total angular momentum of the pion-nucleon system, $j
=l \pm 1/2$. They parametrize the structure of the nucleon as probed with
low energy photons.
Let us first concentrate on the photoproduction case. At threshold,
the differential cross section takes the form
$$ {| \vec k| \over | \vec q | } {d \sigma \over d \Omega} \bigg|_{\vec q
\to 0} = (E_{0+})^2
\eqno(2.11)$$
i.e. it is entirely given in terms of the electric dipole amplitude $E_{0+}$
(which at threshold is real). The quark mass expansion of $E_{0+}$ is given by
$$E_{0+} = {-i e \over 8 \pi m} { \mu ( 2 + \mu ) \over (1 + \mu)^{3/2}}
f({\bar m} / \Lambda_{\rm QCD} ) \eqno(2.12)$$
with $\mu = M_\pi / m \simeq 1/7$ and $\bar m$
the renormalization group
invariant quark masses.  Modulo logarithms, one can expand $f$ in powers of
$\mu$,
$$f = {1 \over \mu} f_{-1} + f_0 + \mu f_1 + {\cal O}(\mu^2 ) \quad .
\eqno(2.13)$$
For charged pions $f_{-1}$ is non-vanishing and one is lead to the famous
Kroll-Rudermann result [20] which states that $E_{0+}$ for $\gamma p \to \pi^+
n$ and $\gamma n \to \pi^- p$ can be expressed in terms of the strong
pion-nucleon coupling constant, the nucleon mass and some kinematical factors.
Consequently, in these cases $E_{0+}$ is finite in the chiral limit. Matters
are different for the case of neutral pions. Here, $f_{-1}$ vanishes and thus
the leading term in the expansion of $E_{0+}$ goes like $\mu$ ($\mu^2$) for
$\gamma p \to \pi^0 p$ ($\gamma n \to \pi^0 n$). The amplitude is therefore
sensitive to the chiral symmetry breaking which makes it a good object to
study experimentally and theoretically. If one performs a one loop CHPT
calculation, one can get unambigously the next-to-leading order terms in the
quark mass expansion of the eletric dipole amplitude. For the production of
the proton, the threshold value is [21]
$$E_{0+} = -{e g_{\pi N} \over 8 \pi m} \mu \biggl\lbrace 1 - \bigl[ {1 \over
2} (3 + \kappa_p ) + ({m \over 4 F_\pi})^2 \bigr] \mu + {\cal O}(\mu^2 )
\biggr\rbrace \quad ,
\eqno(2.14)$$
with $\kappa_p = 3.71$ the anomalous magnetic moment of the proton.
The second term in the square brackets is the new one found in the
CHPT calculation. It stems from the so-called triangle diagram
and its crossed partner. If one expresses $E_{0+}$ in terms of the
conventional Lorentz invariant functions $A_{1,2,3,4}$, one finds that
these diagrams give a contribution $\delta A_1 = ( e g_{\pi N} / 32
F_\pi^2 ) \, \mu$, i.e. they are non-analytic in the quark masses
since $\mu \sim \sqrt{\hat m}$. Remember also that the electric dipole
amplitude at threshold is proportional to $\mu A_1$ and therefore one
gets this novel contribution at next-to-leading order in the quark mass
expansion. Clearly, the expansion in $\mu$ is slowly converging, the
coefficient of the term of order $\mu^2$ is so large that it compensates
the leading term proportional to $\mu$. Therefore, for  a meaningful
prediction one has to go further in the expansion. This has been done
in the relativistic formalism which sums up some of the higher order
corrections [22]. One finds (at threshold) $E_{0+} = -1.33 \pm 0.09$
using the conventional units of $10^{-3} / M_{\pi^+}$. The error
reflects solely the uncertainty in estimating the finite contact term
contributing at order $q^3$.
This is somewhat
below the generally accepted experimental value of $-2.0 \pm 0.2$
 (see e.g.  Bernstein and Holstein [23] or Drechsel and Tiator [24]
and references therein).
However, one should point out that the calculation was performed in the
isospin limit with $M_{\pi^\pm} = M_{\pi^0}$ and $m_p = m_n$. It is a
very tough problem to include isospin-breaking in a systematic fashion
- I invite you to solve it. To get an idea about the effects related
to it, one can perform a toy calculation and set the pion and nucleon
masses by hand on their physical values in the various diagrams. Then,
one finds $E_{0+} = -1.97$, which is an encouraging number but it has
to be confirmed by a better calculation. The chief reason for the
difference to the isospin-symmetric case is that the contribution of the
triangle diagram now vanishes below the $\pi^+ n$ threshold. It is
furthermore interesting to observe that the one-loop corrections to the
electric dipole amplitudes of charged pion photoproduction move the
prediction closer to the data. These chiral corrections are suppressed
by powers of $\mu^2$ and $\mu^2 \ln \mu$ with respect to the leading
Kroll-Ruderman term ($\sim \mu^0$) and therefore a more accurate
experimental determination of these quantities is called for.
Let me now consider the electroproduction process. First, the low-energy
 theorems for the two $S$-wave multipoles $E_{0+}$ and $L_{0+}$ (for
virtual photons, there is also a longitudinal coupling to the nucleon)
 have
 been discussed in the light of CHPT in ref.[25]. Let me stress that
 the LETs derived in refs.[21,25] are the ones implied by QCD and that
 they can be tested experimentally. Notice that there still seems to
 be some confusion concerning the meaning and the
 interpretation of these theorems [26]. However,
 I will not dwell on this topic here.
 Instead, I want to direct your
 attention to some new data. Welch et al.[27] have published the
 $S$-wave cross section for the reaction $\gamma^\star p \to \pi^0 p$
 very close to the photon point. This measurement is a quantum step
 compared to previous determinations which mostly date back to the
 seventies when pion electroproduction was still a hot topic in particle
 physics. In this experiment, $k^2$ varied between -0.04 and -0.10
 GeV$^{-2}$ and the $S$-wave cross section could be extracted with
 an unprecedented accuracy (see fig.2 in [27]). This is also the
 kinematical regime where a CHPT calculation might offer some insight.
 Indeed, in ref.[28] it was shown that  the $k^2$-dependence of this
 cross section seems to indicate the necessity of loop effects. With
 conventional models including e.g. form factors and the anomalous
 magnetic moment coupling the trend of the data can not be described.
 However, the corrections from the one loop diagrams to the tree level
 prediction are substantial. This gives further credit to the previously
 made
 statement that a calculation beyond next-to-leading order should be
 performed. The last topic I want to address in this section concerns
 the determination of the nucleon  axial radius from charged pion
 electroproduction. Let me briefly explain how the axial form factor
 comes into play. The basic matrix element to be considered is the
 time-ordered product of the electromagnetic (vector)  current with
 the interpolating pion field sandwiched between nucleon states. Now
 one can use the PCAC relation and express the pion field in terms of
 the divergence of the axial current. Thus, a commutator of the form
[V,A] arises. Current algebra tells us that
this gives an axial current between the incoming and outgoing
nucleon fields and, alas, the axial form factor. The isospin factors
combine in a way that they form a totally antisymmetric combination
which can not be probed in neutral pion production. These ideas were
formalized in the venerable low-energy theorem (LET)
due to Nambu, Luri{\'e}
and Shrauner [29] for the isospin--odd electric dipole amplitude
$E_{0+}^{(-)}$ in the chiral limit,
$$
E_{0+}^{(-)}(M_\pi=0, k^2) ={e g_A \over 8 \pi
F_\pi} \biggl\lbrace 1 +{k^2 \over 6} r_A^2 + { k^2 \over 4m^2} (\kappa_V
+ {1 \over 2}) + {\cal O} (k^3) \biggr\rbrace
\eqno(2.15)$$
Therefore, measuring the reactions
$\gamma^\star p \to \pi^+ n$ and
$\gamma^\star n \to \pi^- p$ allows to extract
$E_{0+}^{(-)}$ and one can determine the axial radius of the nucleon, $r_A$.
This quantity measures the distribution of spin and isospin in the nucleon,
i.e. probes the Gamov--Teller operator $\bf  \sigma \cdot  \tau$. A
priori, the axial radius is expected to be different from the typical
electromagnetic size,
$r_{{\rm em}} \simeq 0.85$ fm. It is customary to parametrize
the axial form factor $G_A (k^2)$ by a dipole form, $G_A (k^2) = (1 - k^2 /
M_A^2 )^{-2}$ which leads to the relation $r_A = \sqrt{12} / M_A$. The axial
radius determined from electroproduction data is typically $r_A = 0.59 \pm
0.04$ fm
whereas (anti)neutrino-nucleon reactions lead to somewhat
larger values, $r_A = 0.65 \pm 0.03$ fm. This discrepancy is usually not
taken seriously since the values overlap within the error bars. However, it
was shown in ref.[30] that pion loops modify the LET (2.15) at order $k^2$
for finite pion mass. In the heavy mass formalism, the coefficient of the
$k^2$ term reads
$$ {1 \over 6} r_A^2 + {1 \over 4 m^2}(\kappa_V +{1 \over2}) +
{1 \over 128 F_\pi^2} (1 - {12 \over \pi^2})
\eqno(2.16)$$
where the last term in (2.16) is the new one. This means that previously one
had extracted a modified radius, the correction being $3 (1 - 12/\pi^2 ) / 64
F_\pi^2 \simeq -0.046$ fm$^2$. This closes the gap between the values of
$r_A$ extracted from electroproduction and neutrino data. It remains to be
seen how the $1 / m$ suppressed terms will modify the result (2.16). Such
investigations are underway.
\titleb{2.5}{Nucleon Compton Scattering}
Consider low-energy (real) photons scattering off a spin-1/2 particle. In
forward direction, the scattering amplitude takes the form
  $$T_{1/2}(\omega) = f_1(\omega^2)\,\,{\bf  \epsilon}_f^* \cdot {\bf
 \epsilon}_i
  + i \omega \, f_2(\omega^2)\,\, {\bf \sigma}\cdot ({\bf \epsilon}_f^* \times
  {\bf \epsilon}_i)
\eqno(2.17)$$
where $\omega$ is the photon frequency and the ${\bf \epsilon}_{i,f}$ are the
polarization vectors in the initial and final state, respectively. The energy
expansion of the spin-dependent amplitude $f_1 (\omega^2 )$ reads
  $$f_1(\omega^2) = - {e^2Z^2\over 4 \pi       m} + (\alpha + \beta) \,\omega^2
 +
  {\cal O}(\omega^4)
\eqno(2.18)$$
where the first energy-independent term is nothing but the Thomson amplitude
mandated by gauge invariance. Therefore, to leading order, the photon  only
probes some global properties like the mass or electric charge of the spin-1/2
target. At next-to-leading order, the non-perturbative structure is
parametrized by two constants, the so-called electric and magnetic
polarizabilities (more correctly, one might want to talk about "magnetic
susceptibility", however, I will use the common language). A detailed
discussion of the CHPT calculation of these fundamental two-photon observables
is given in ref.[3] (see also the references quoted there). Here, I just want
to add a few educational remarks. For the proton, the Thomson term is
non-vanishing and it is interesting to compare its magnitude to the one from
the polarizability contribution using $(\alpha + \beta)_p = 14.2 \cdot
10^{-4}$
fm$^3$. The recent Illinois Compton scattering experiment [31] to determine
$\alpha_p$ and $\beta_p$ was performed at photon energies ranging from 32 to
72 MeV. While at the lowest $\omega$ the polarizability term amounts to a 31
per cent correction to the Thomson term contribution ($= -1.22 \cdot 10^{-4}$
fm),
at 72 MeV the second term in the energy expansion is 1.5 times as large as the
leading term.  Clearly, higher order ($\omega^4$) contributions have to be
taken into account at such energies. Another remark concerns the behaviour of
$\alpha_{p,n}$ and $\beta_{p,n}$ in the chiral limit.  These quantities diverge
as 1/$M_\pi$ as the pion mass tends to zero. This is expected since the two
photons probe the long-ranged pion cloud, i.e. there is no more Yukawa
suppression as in the case for a finite pion mass. Now a well-known  dispersion
sum rule relates $(\alpha + \beta)$ to the total nucleon
photoabsorption cross section.
The latter is, of course, also well-behaved in the chiral limit which at first
sight seems to be at variance with the behaviour of the expansion of the
scattering amplitude. But be aware that the general form of (2.18) has been
derived under the assumption that there is a well defined low-energy limit.
Similar
observations can also be made concerning the chiral expansion of the $\pi \pi$
scattering amplitude discussed in section 2. Let me now focus on the
spin-dependent amplitude $f_2 (\omega^2 )$ which has an expansion analogous to
(2.18),
  $$f_2(\omega^2) = f_2(0) + \gamma\, \omega^2 + {\cal O}(\omega^4)
\eqno(2.19)$$
By the optical theorem, the imaginary part of $f_2$ is related to the
photoabsorption cross sections for circularly polarized photons on polarized
nucleons,
$${\rm Im} \, f_2 (\omega^2 ) = - {1 \over 8 \pi} \biggl[ \sigma_{3/2} (\omega
) - \sigma_{1/2}(\omega ) \biggr] \eqno(2.20)$$
where the indices 3/2 and 1/2 refer to the total $\gamma N$ helicities. We will
come back to this relation later on. Let me now turn to the discussion of the
Taylor coefficients $f_2 (0)$ and $\gamma$ in eq.(2.19). There exists a
celebrated LET for $f_2 (0)$ due to Low, Gell-Mann  and Goldberger [32]. Using
very general principles like gauge invariance, Lorentz invariance and crossing
symmetry, they showed that
$$f_2 (0) = -{e^2 \kappa^2 \over 8 \pi m^2} \, \, . \eqno (2.21)$$
Here, $\kappa$ denotes the anomalous magnetic moment of the particle the
photon scatters
off. In CHPT, one can easily derive the LET making use of the heavy baryon
formalism as shown in ref.[33]. It stems from some contact terms which are
nominally suppressed by powers of 1/m. To one-loop order, one can give $f_2
(\omega^2 )$ in closed form since only a few loop diagrams contribute (in
addition to the contact terms giving the LET),
  $$f_2(\omega^2) = -{e^2 \kappa^2 \over 8 \pi m^2} + {e^2
 g_A^2
  \over 32 \pi^3 F^2 } \biggl[ {M^2 \over \omega^2} \, \arcsin^2 ({\omega\over
M})   - 1 \biggr]
\eqno(2.22)$$
{}From this the slope parameter $\gamma$ follows immediately. Notice that
$\gamma$ is again the sum of an electric and magnetic piece, but in the
absence of any data, I will only discuss the sum. The em decomposition can be
supplied upon request. To leading order, $\gamma$ is identical for the p and
the n,
$$\gamma = {e^2 g_A^2 \over 96 \pi^3 F_\pi^2 M_\pi^2} = 4.4 \cdot 10^{-4} \,
{\rm fm}^4 \, \, \, . \eqno(2.23)$$
In ref.[33] it was shown that including 1/m suppressed effects in the
relativistic formalism together with an estimate of the $\Delta (1232)$
resonance contribution leads to $\gamma_p = -1.5$ and $\gamma_n = -0.5$ (in
conventional units), in fair agreement with the estimate based on the
dispersion sum rule
$$\gamma = - {1 \over 4 \pi^2} \int_{M_\pi}^\infty {d\omega \over \omega^2}
\biggl[ \sigma_{3/2} (\omega ) - \sigma_{1/2}(\omega ) \biggr] \eqno(2.24)$$
which follows directly from eq.(2.20). Using the helicity cross sections
provided by Workman and Arndt [34] one gets $\gamma_p^{{\rm emp}} = -0.89$ and
$\gamma_n^{{\rm emp}} = -0.53$. Clearly, a direct experimental determination
of these observables is called for. To give an idea about the size of the
effect induced by the "spin-polarizability" $\gamma$, let us compare the
leading order term $f_2 (0) = -5.19 \cdot 10^{-4}$ fm$^2$ with the correction
from $\gamma \omega^2$ at various energies. For $\omega = 50$ MeV, this term
amounts to a 6 per cent reduction of the leading one. At 100 MeV photon
energy, the correction has grown to 22 per cent. For comparison, evaluating
the full one-loop amplitude (2.22), one finds corrections of 7 and 37 percent
for $\omega = 50$ and 100 MeV, respectively. These latter numbers will, of
course, be changed by higher loop corrections. We notice that the influence of
the next-to-leading order term is much less pronounced than in the case of the
spin-independent amplitude and thus an accurate determination of $\gamma$ and
its electric and magnetic components will be more difficult. The last topic I
want to address in this section is the so-called Drell-Hearn-Gerasimov [35]
sum rule and its extension to virtual photons. For photons with four-momentum
$k^2 \le 0$ the extended DHG sum rule takes the form
$$I(k^2) =  \int_{\omega_{{\rm thr}}}^\infty {d\omega \over \omega  }
\biggl[ \sigma_{1/2} (\omega , k^2) - \sigma_{3/2}(\omega , k^2 ) \biggr]
\eqno(2.25)$$
For real photons, this reduces to the DHG sum rule, which has not yet been
tested experimentally. For the proton, one has $I(0) =
- \pi e^2 \kappa_p^2 /
2 m^2 = -0.53$ GeV$^{-2}$. At large $k^2 \simeq - 10$ GeV$^2$, the EMC [36]
data tell  us that $I$ is positive and thus has to undergo a sign change. CHPT
allows one to calculate the slope of $I(k^2)$ in the vicinity of the photon
point [37]. While the most recent phenomenological analysis gives a kink
structure at small $k^2$ [38], in CHPT one finds that $I(k^2)$ increases
monotonically. At present, one can not decide upon this fine detail but has to
wait for the experimental determinations which are possible with CW
machines and are planned (MAMI, CEBAF, $\ldots$). Finally, I wish to stress
again that CHPT allows to systematically explore the consequences of the
spontaneously broken chiral symmetry of QCD. It is {\bf not} a model as should
have become clear from the above discussion but rather an exact
non-perturbative method.
\titlea{3}{The Quark Structure of the Nucleon}
In this section I will be concerned with the quark structure of the nucleon as
revealed from measurements using electroweak probes. In particular, one can
address the questions surrounding the possible admixtures of strange quark
components into the proton's wave function. Since these are mostly small
effects, I will also discuss the effects of heavy quark and radiative
corrections to the processes which allow one to determine the electroweak form
factors.
\titleb{3.1}{Currents, Sizes and Form Factors of the Nucleon}
The starting point of the discussion are the well-known electromagnetic and
axial currents based on photon and charged vector boson exchanges. Let us
first
consider the vector current $J_\mu$ stemming from the one-photon exchange.
Between nucleon states of momentum $p$ and helicity $s$ (in general, I will
suppress helicitiy indices) Lorentz-invariance, parity and charge conjugation
allow us to write
$$< N(p',s') | J_\mu | N(p,s)> = e {\bar u}(p',s') \bigl[ F_1 (q^2) \gamma_\mu
+ F_2 (q^2) {i \sigma_{\mu \nu} q^\nu \over 2 m} \bigr] u(p,s) \eqno(3.1)$$
with $q_\mu= (p'-p)_\mu$ the four-momentum transfer, $m$ denotes the nucleon
mass and $u(p,s)$ a conventional nucleon spinor. The finite extension
(non-perturbative structure) is parametrized in terms of the Dirac and Pauli
form factors (ffs), $F_1 (q^2)$ and $F_2 (q^2)$, respectively. Gauge
invariance of the electromagnetic interactions demands $\partial_\mu J^\mu (x)
= 0$. The physical interpretation of these ffs is particularly transparent in
the Breit (brick wall) frame, in which the photon transfers no energy, $q_\mu
= (0, \vec q )$. The time component of $J_\mu$ is the charge density, $J_0 =
\rho$ and the space components lead to the distribution of magnetism,
$$\eqalign{
< N (\vec q /2,s') | J_0 | N (-\vec q /2,s)> &= e G_E (q^2) \delta_{s's} \cr
\, \, &= e [F_1 (q^2) - \tau F_2 (q^2) ] \delta_{s's} \cr
< N (\vec q /2,s') | \vec J | (-\vec q /2,s)> &= {e \over 2m} G_M (q^2)
\chi_{s'}^\dagger i {\bf  \sigma} \times {\vec q} \chi_s \cr &= {e\over 2m}
[F_1 (q^2)
+ F_2 (q^2) ] \chi_{s'}^\dagger i {\bf \sigma} \times {\vec q} \chi_s \cr}
\eqno(3.2)$$
with $\chi_s$ a two-component spinor and $\tau = -q^2 / 4m^2 = Q^2 / 4m^2$
($Q^2 > 0$ for space-like photons). Fourier-transformation of the electric
charge density leads to
$$\eqalign{ \int \rho({\vec r})j_0 (qr) d^3 r &= \int \rho({\vec r})d^3 r -
{1 \over 6} Q^2 \int \rho({\vec r})r^2 d^3 r + \ldots \cr &= Z - {1 \over 6}
Q^2 <r^2>_E + \ldots \cr} \eqno(3.3)$$
where the first term is obviously the electric charge of the particle (here,
the proton or the neutron) and the second one defines the electric charge
radius. It is conventional to normalize these charge radii via
$$<r^2 > = {6 \over G(0)} {dG(q^2) \over dq^2} \bigg|_{q^2 = 0} \eqno(3.4)$$
with the exception of the neutron charge radius  since in that case $G(0) = 0$
so that $<r^2_E >_n$ is simply six times the slope of the neutron form factor.
The electric (E) and magnetic (M) ffs are normalized as follows
$$G_E^p (0) = 1, \quad G_E^n (0) = 0, \quad G_M^p (0) = \mu_p = 2.793 , \quad
G_M^n (0) = \mu_n = -1.913 \,  \, \, . \eqno(3.5a)$$
For the various radii, the presently available data are [39]
$$\eqalign{ <r_E^2>_p^{1/2} = \, \, &0.86 \pm 0.01 \, {\rm fm}, \quad
<r_M^2>_p^{1/2} = 0.86 \pm 0.06 \, {\rm fm}, \cr
<r_E^2>_n
= -&0.12 \pm 0.01 \, {\rm fm}^2, \quad
<r_M^2>_n^{1/2} = 0.88 \pm 0.07 \, {\rm fm} \, \, . \cr} \eqno(3.5b)$$
A few remarks on these numbers are in order. First, the typical
electromagnetic (em) size is $r_{\rm em} \simeq 0.85$ fm and the negative
value of the neutron charge radius squared shows that a surplus of negative
charge in the
exterior region of the neutron must exist.  Most disturbing is the fact that
in particular the magnetic charge radius of the neutron, which is a quantity
as fundamental as say the magnetic moment, is only poorly known. As pointed
out by Arenh\"ovel and Schoch at this school [40] experiments are under way to
remedy the situation. At larger momentum transfers, it has become common to
use the so-called dipole fits, which proved to be embarrasingly good up to
several GeV$^2$ on typical logarithmic plots when elastic electron-proton
scattering experiments were
performed in the sixties and early seventies as discussed by Taylor here [41].
The dipole fit reads
$$G_E^p (Q^2) = {G_M^p (Q^2) \over \mu_p} = {G_M^n (Q^2) \over \mu_n} = G_D
(Q^2) = {1 \over (1+Q^2/0.71 \,{\rm GeV^2})^2} \, \, . \eqno(3.6)$$
For the electric ff of the neutron, the "trivial" dipole fit is $G_E^n
(Q^2)=0$, however, in nature this ff is small but non-vanishing. It can be
approximated by [42]
$$G_E^n (Q^2) = - {\mu_n \tau \over 1 + Q^2/ 0.42 \, {\rm GeV^2}} G_D (Q^2) \,
\, . \eqno(3.7)$$
In refs.[43], you can take a look at the quality of these fits on a linear
scale for $Q^2$ up to 4 GeV$^2$ - it is obvious that the dipole fit is at most
a fair approximation. Also, it has never been given a sound theoretical
foundation. While the dipole mass $M_D = 0.843$ GeV is suggestive of vector
meson exchange (like in the vector meson dominance approach to photon-hadron
couplings), such a mechanism would naturally lead to a monopole fall-off. In
case of the neutron electric ff, which is determined indirectly from elastic
scattering off the deuteron and subtraction of the proton ff, a major
uncertainty stems from the use of the underlying model of the nucleon-nucleon
force as discussed in ref.[42]. Here, experiments using polarized electrons
and polarized helium targets are hoped to improve the theoretical
uncertainties. Definitively, we need a better determination of these
fundamental nucleon properties. It is also instructive to see what
perturbative QCD tells us about the large $Q^2$ behaviour  of these ffs. Based
on the quark counting rules [44], which state that to distribute the large
photon momentum equally to three quarks in the nucleon, two gluon exchanges
are necessary, one finds that $F_1 (Q^2) \sim 1/Q^4$ and $F_2 (Q^2) \sim
1/Q^6$
since the gluon propagator goes like $1/Q^2$ and in the case of $F_2$
an extra $1/Q^2$ is needed for the helicity flip. At which value of $Q^2$ this
asymptotic behaviour sets in is not known, certainly way beyond 1 GeV$^2$ as
originaly thought. Notice that the dipole fits to the nucleon ffs have the
correct large $Q^2$ behaviour. This is a further mystery surrounding these
fits. It might be based on the fact that a simple meson-cloud model of the
nucleon also leads to such fall-offs (in the one-boson exchange
approximation). Let me now switch to the axial current (related to the
exchange of charged vector bosons). Its nucleon matrix element reads
$$<N(p') |A_i^\mu (0) | N(p)> = \bar u(p') \bigl[ G_A (q^2) \gamma^\mu + {G_P
(q^2) \over 2 m} q^\mu \bigr] \gamma_5 {\tau_i \over 2} u(p) \eqno(3.8)$$
where $G_A (q^2)$ is the axial ff.  The induced pseudoscalar ff $G_P (q^2)$ is
essentially pole dominated and only accessible via scattering processes
involving heavy leptons (like e.g. muons). I will not consider it in what
follows. $G_A (0)$ is nothing but the axial-vector coupling constant measured
in neutron $\beta$-decay, $g_A = 1.26$. The $Q^2$-dependence of $G_A (Q^2)$
follows again a dipole form,
$$ G_A (Q^2) = {g_A \over (1+Q^2/M_A^2)^2} \quad. \eqno(3.9)$$
The value of the axial cut off mass was already discussed in section 2.4, it
translates into a typical axial size of $r_A \simeq 0.65$ fm. This is
considerably smaller than the typical em size. Therefore, the size of the
nucleon depends on the probe one uses. In fact, the hierachy of these various
nucleon sizes can nicely be understood in the topological soliton model of
the nucleon as spelled out in detail in ref.[45].
In a nutshell, the argument goes as follows. The isoscalar baryon number
current, whose space-integral is the baryon number, is given by the pion
fields in the soliton with an extension of $r_B = 0.5$ fm. The isoscalar
photon, however, sees in addition the virtual $\omega$-meson content of the
nucleon. This leads to an isoscalar charge radius of $<r_E^2>_{I=0} = r_B^2 +
6/M^2_{\omega} = (0.8$ fm)$^2$ in nice agreement with the data. By a similar
argument, one expects a somewhat smaller axial radius since the additional
factor now carries the axial vector mass of 1.1 GeV and one is thus naturally
lead to an axial extension of 0.65 fm.

\titleb{3.2}{A Toy Model: Quark Distributions from Nucleon Form Factors}

To understand how the diferent electroweak properties as parametrized by the
various ffs can give insight into the quark structure of the nucleon in the
non-perturbative regime, let us consider a simple (too simple) model. The
proton and the neutron are made up of valence u and d quarks and a sea of
quark-antiquark pairs. We assume that the p and the n are simply related by the
interchange of one valence u with one valence d quark.
For the sake of
simplicity, let us forget the gluons and just imagine some confinement
mechanism
which keeps the current quarks within the typical hadronic volume of 1 fm$^3$.
Denoting by $J_\mu^j$ the em current associated to the quarks of flavor $j =
u,d,s,c,b,t$, we can write [46]
$$\eqalign{<N(p',s') | J_\mu | N(p,s)> &= <N(p',s') | \sum_j J_\mu^j | N(p,s)>
\cr &= e \sum_j {\bar u}(p',s') Q_j \bigl( \gamma_\mu F_1^j + { i \sigma_{\mu
\nu} q^\nu \over 2m } F_2^j \bigr) u(p,s) \cr} \eqno(3.10)$$
where $Q_j$ denotes the quark charges and the quarks are considered as
point-like Dirac particles with no anomalous magnetic moment. In terms of the
electric and magnetic ffs, this decomposition takes the form
$$G_{E,M}^{(p,n)} = \sum_j Q_j G_{E,M}^{j \,(p,n)} \quad . \eqno(3.11)$$
These expressions are exact under the assumptions made as long as the sum
extends over all quark flavors. Furthermore, the $G^j$ contain contributions
from quarks as well as from antiquarks. Therefore, the $q \bar q$ sea does not
contribute to the total charge since $Q_j = - {\bar Q}_j$. Since the p and the
n as well as the u and the d quark are considered to be essentially the same
particles differing only by an isospin rotation, it should be possible to
separate the response of the various quark flavors [47]. For doing that, one
has to make further assumptions. First, one postulates a local isospin
invariance of the type $J_\mu^{u,p} (x) = J_\mu^{d,n} (x)$ and
$J_\mu^{d,p} (x) = J_\mu^{u,n} (x)$ (which has never been  tested and is
doubtful in the light of the recent structure function measurements). Second,
we assume that all other quark flavors do not contribute, i.e. $J_\mu^{q,p}(x)
=
J_\mu^{q,n}(x) = 0$ for $q = s,c,b,t$. This allows to uniquely give the em u
and d
distributions in terms of appropriate combinations of the em ffs of the proton
and
the neutron. The final formulae are obtained by normalizing with respect to the
total quark charges, e.g. pulling out a factor 4/3 for the u quark distribution
in the proton. This leads to [48]
$$\eqalign{ G^u_{E,M} &= G^p_{E,M} + {1 \over 2} G^n_{E,M} \, \, ,  \cr
G^d_{E,M} &= G^p_{E,M} + 2 G^n_{E,M} \, \, . \cr} \eqno(3.12)$$
It is instructive to get some numbers from eq.(3.12). First, let us stick in
the magnetic moments. We find $G_M^u (0) = 1.836$ and $G_M^d (0) = -1.033$,
which is not very different from the $SU(3)$ result for constituent quarks.
Notice, however, that here we are considering current quarks. The implications
of this result are discussed in more detail in Beck's lecture [48]. Similarly,
for the electric charge radii one finds $<r^2_E>_u = 0.68 \pm 0.02$ fm$^2$
and $<r^2_E>_d = 0.51 \pm 0.02$ fm$^2$ which is a 35 per cent difference.  To
further separate the valence from the sea quarks, one needs additional
assumptions about the radial distributions as detailed in ref.[48]. In any
case, this discussion should only be considered illustrative, the model is too
crude to attach a deep physical significance to these numbers. However, in a
very simple fashion it demonstrates how a clever combination of measurable
hadron form factors allows us to make statements about quark distributions.
This problem will now be tackled in a more serious fashion.
\titleb{3.3}{Electroweak Currents}
In this section, I will define the electroweak
currents related to the
exchanges of the various vector bosons of the minimal standard model. These
spin-1 particles are the massless photon and the massive $W^\pm$ and $Z^0$
bosons. At tree level, the couplings of these to the various quark flavors are
specified completely in terms of the $U(1)$ gauge coupling $g$ and the $SU(2)$
gauge coupling $g'$. Equivalently, one can use the em coupling $e$ and weak
mixing angle $\theta_W$ since $g = e \sin \theta_W = e \, s$ and
$g' = g / 4 \cos \theta_W = g /4 \, c$. Also, the vector boson masses are
related
at tree level via $M_Z = M_W / c$. For the most recent values, see Altarelli's
lectures at this school [49]. Let us consider again the photon current which
is purely vectorial and its couplings to the light quarks u, d and s. One has
$$\eqalign{ J_\mu^\gamma &= \sum_{q=u,d,s} Q_q {\bar q} \gamma_\mu q =
{2 \over 3} ( {\bar u} \gamma_\mu u) - {1 \over 3} ({\bar d} \gamma_\mu d +
{\bar s} \gamma_\mu s) \cr
&= \sum_{q=u,d,s} {\bar q} \gamma_\mu {1 \over 2} \bigl( \lambda^3 + {1 \over
\sqrt{3}} \lambda^8 \bigr) q \cr
&= {1 \over 2} ( {\bar u} \gamma_\mu u - {\bar d} \gamma_\mu d ) +
{1 \over 6} ( {\bar u} \gamma_\mu u + {\bar d} \gamma_\mu d - 2 {\bar s}
\gamma_\mu s ) \quad . \cr} \eqno(3.13)$$
I have exhibited the two most frequently used conventions. In the second line,
the decomposition in terms of Gell-Mann's flavor $SU(3)$ matrices is shown.
One notices that the photon is blind to the singlet piece $\sim \lambda^0$. In
the third line, the photon current is decomposed into its isovector (IV) and
isoscalar (IS) components. This current leads to the em form factors
$G_{E,M}^\gamma$ or $F_{1,2}^\gamma$ discussed before (the label '$\gamma$' is
needed to differentiate these ffs from the ones related to the weak neutral
current as discussed below). The charged weak currents are flavor-changing and
lead to the isovector axial ff discussed before. Further information comes
from the weak neutral current mediated by the $Z^0$. Its coupling to the quark
has a vectorial and an axial-vector piece. Ignoring the overall coupling
strength $g'$ and defining $V_q = {\bar q} \gamma_\mu q$ for a quark of flavor
$q$, the vector part at tree level reads
$$\eqalign{ V_\mu^Z &= \bigl( {1 \over 4} - {2 \over 3} s^2 \bigr) V_u +
\bigl( -{1 \over 4} + {1 \over 3} s^2 \bigr) (V_d + V_s ) \cr
&= {1 \over 2} \bigl( {1 \over 2} - s^2 \bigr) (V_u - V_d ) +
{1 \over 6} \bigl( {1 \over 2} - s^2 \bigr) (V_u + V_d - 2 V_s ) - {1 \over
12} (V_u + V_d + V_s ) \cr
&= \bigl( {1 \over 2} - s^2 \bigr) {1 \over 2} (V_u - V_d ) +  {1 \over 6} s^2
(V_u + V_d - 2 V_s ) - {1 \over 4} V_s \cr} \eqno(3.14)$$
where again the second line gives the $SU(3)$ components and the third line
the decomposition into IS, IV and explicit strange pieces. Similarly, with
$A_q = {\bar q} \gamma_\mu \gamma_5 q$ we have for the axial-vector current
$$\eqalign{ A_\mu^Z &= -{1 \over 4} A_u + {1 \over 4} (A_d + A_s ) \cr
&= -{1 \over 4} (A_u - A_d) - {1 \over 12} (A_u + A_d - 2 A_s ) +
{1 \over 12} (A_u + A_d + A_s ) \cr
&= -{1 \over 2} \bigr[ {1 \over 2} (A_u - A_d) \bigr] + {1 \over 4} A_s \cr}
\eqno(3.15)$$
Two remarks are in order. First, the $Z^0$ couples to the singlet component
$\sim \lambda^0 \sim (u+d+s)$ and, second, there is no isoscalar axial
coupling at tree level. For the following discussion, let me express the ffs
related to the $Z^0$-exchange in the flavor $SU(3)$ basis following Kaplan and
Manohar [50],
$$\eqalign{F_{1,2}^Z &= \sum_{\alpha = 0,3,8} (b_\alpha - a_\alpha s^2 )
F^\alpha_{1,2} \, ; \quad F_{1,2}^\gamma = \sum_{\alpha = 0,3,8} a_\alpha
F^\alpha_{1,2} \, \, ,  \cr G_1 &= - \sum_{\alpha
= 0,3,8} b_\alpha G_A^\alpha = - \sum_{\alpha = 0,3,8} b_\alpha G_1^\alpha
\, \, , \cr a^0 &= 0 \, , \quad a^3 = 1 \, , \quad a^8 = {1 \over \sqrt{3}} \,
; \quad b^0 = -{1 \over 4} \, , \quad b^3 = {1 \over 2} \, , \quad b^8 = {1
\over 2 \sqrt{3}} \, . \cr} \eqno(3.16)$$
The value of the a's and b's are subject to radiative and heavy quark
corrections as discussed below. For completeness, I have also given the em ffs
in this basis. Using for the $SU(3)$ generators Tr($T^a T^b)
= \delta^{ab}/2$ and supplying the isosinglet currents with an overall factor
1/3, the normalizations of the various ffs are given as follows. From the
proton and the neutron charge, we have
$$F_1^3 (0) = 1/2 \, , \quad F_1^8 (0) = \sqrt{3} / 2 \quad . \eqno(3.16a)$$
Similarly, the p and n anomalous magnetic moments lead to
$$F_2^3 (0) = (\kappa_p - \kappa_n) /2 \, , \quad F_2^8 (0) = \sqrt{3}
(\kappa_p + \kappa_n) / 2 \quad . \eqno(3.16b)$$
$F_1^0 (0)$ is equal to unity since it gives the baryon number. However, this
does not exclude a finite extension of $s \bar s$ pairs leading to a strange
electric radius (in complete analogy to the neutron charge radius). The value
of the singlet anomalous magnetic moment is not known. If there are no strange
quarks in the proton, one has $F_2^0 (0) = \kappa_p + \kappa_n = -0.12$. So a
measurement of $F_2^0 (0)$ will give information about the possible strange
quark contribution to the anomalous magnetic moment,
$F_2^0 (0) = \kappa_p + \kappa_n + \mu_s$. The triplet and octet
normalizations of the axial ff are given by
$$G_1^3 (0) = g_A = F + D = 1.26 \, , \quad
G_1^8 (0) = ( 3 F - D )/ \sqrt{3} = 0.32 \, . \eqno(3.16c)$$
The isosinglet charge
$G_1^0 (0)$ is again sensitive to the strange quark
content and can e.g. be determined in polarized deep inelastic lepton
scattering as discussed by Reya here [51] (see also ref.[36]). Alternatively,
neutrino scattering
off nucleons or nuclei might be used to get a handle on $G_1^0 (0)$. This will
be discussed below. At present, no experimental information is available about
the $Q^2$-dependence of the singlet and octet axial ffs, so one either has to
resort to some assumptions or models (see section 3.5). Clearly, the singlet
axial charge and the singlet anomalous magnetic moment are the best objects to
study the strange quark content of the nucleon using electroweak probes. The
other obvious candidate is the so-called pion-nucleon $\sigma$-term which
measures the strength of the strange matrix element $m_s <p|{\bar s}s|p>$.
Combining $\pi N$ data, dispersion theory and constraints from chiral
symmetry, Gasser et al. [52] have given the most precise determination of this
quantity and its $t$-dependence. From this analysis one deduces that
$m_s <p|{\bar s}s|p> \simeq 130$ MeV. This is a sizeable though not dramatic
strange quark effect. After defining the electroweak form factors, I will now
discuss some experiments which allow to extract them and the related strange
quark matrix elements.
\titleb{3.4}{Parity-Violating Electron Scattering}
Consider the scattering of polarized electrons on a nucleon or a nucleus. Due
to the $\gamma - Z^0$ interference, parity violation (pv) occurs and this
allows to determine combinations of the electromagnetic and weak neutral ffs.
The basic Feynman diagrams are simply the one-photon exchange proprotional to
$F_{1,2}^\gamma$ and the $Z^0$-exchange proportional to $F_{1,2}^Z$ and $G_1$.
To get an idea about the magnitude of this interference effect, let us perform
some dimensional analysis. The elctromagnetic amplitude reads
$${\cal A}_{\rm em} = < f | J^\mu_{\rm em} {1 \over Q^2} j_{\mu,{\rm em}} | i
> \sim {\alpha \over Q^2} \, \, , \eqno(3.17)$$
where $j_{\mu,{\rm em}}$ is the well-known em lepton current, $1/Q^2$ the
photon propagator and $J^\mu_{\rm em}$ the hadronic vector current
parametrized by the em ffs. The parity-violating weak amplitude takes the form
$${\cal A}_{\rm w}^{\rm pv} = < f | V^\mu a_\mu + A^\mu v_\mu | i > \sim
{G_F  \over \sqrt{2} } \, \, , \eqno(3.18)$$
which is a product of the lepton ($v_\mu,a_\mu$) and hadron ($V_\mu,A_\mu$)
vector and axial-vector currents, the latter containing the weak form factors.
$G_F = 1.16 \cdot 10^{-5}$ GeV$^{-2}$ is the Fermi constant. The total cross
section can approximately be written as
$$\sigma = |{\cal A}_{\rm em} + {\cal A}_{\rm w}^{\rm pv}|^2
\simeq |{\cal A}_{\rm em}|^2 +
2 | {\cal A}_{\rm em}^* {\cal A}_{\rm w}^{\rm
pv}| \, \, \, . \eqno(3.19)$$
The observable of interest is the left-right asymmetry $A$,
$$A = {\sigma_R - \sigma_L \over \sigma_R + \sigma_L} \simeq
{ | {\cal A}_{\rm em}^* {\cal A}_{\rm w}^{\rm pv} | \over
|{\cal A}_{\rm em}|^2} \sim {G_F m^2 \over 4 \pi \alpha} \, {Q^2 \over m^2} =
1.1 \cdot 10^{-4} {Q^2 \over {\rm GeV}^2} \, \, . \eqno(3.20)$$
For a typical four-momentum of $Q^2 = 0.1$ GeV$^2$ in the non-perturbative
regime we can expect an asymmetry of 10$^{-5}$. This is a small number but
within the reach of present day technology. Historically, the first
measurement of this asymmetry using a deuterium target at SLAC was performed
at $Q^2$ around 20 GeV$^2$ [53]. To be specific, consider now as the target
the proton. An elementary calculation gives
$$\eqalign{A_{{\vec e}p} &= -{G_F Q^2 \over \sqrt{2} \pi \alpha} {\bigl[
\epsilon G_E^\gamma G_E^Z + \tau G_M^\gamma G_M^Z -{1 \over 2}(1 - 4 s^2)
\sqrt{1- \epsilon^2} \sqrt{\tau (1+\tau)} G_M^\gamma G_1 \bigr] \over
\epsilon (G_E^\gamma )^2 + \tau (G_M^\gamma )^2}  \cr
&= -{G_F Q^2 \over \sqrt{2} \pi \alpha} [A_E + A_M + A_A ] \cr} \eqno(3.21)$$
with $1/\epsilon = [1 + 2(1+\tau) \tan^2 (\theta / 2)] = 0 \ldots 1$ given by
the scattering angle of the electron. In the second line, I have split the
asymmetry into an electric, magnetic and axial piece. It is instructive to
consider various kinematical regimes. At forward angles ($\theta \to 0, \,
\epsilon \to 1$), the electric part dominates at low $Q^2$ and the asymmetry
is
thus sensitive to the strange electric ff. At higher $Q^2$, due to the $\tau$
prefactor, the magnetic part takes over and one can access $G_M^s (Q^2 )$.
Notice that in forward direction  the axial contribution is not only
suppressed by $1 - 4 s^2 \simeq 0.08$ but also by $\sqrt{1-\epsilon^2}$. In
backward direction, matters are different since as $\theta \to 0$, $\epsilon$
also tends to zero. So at low $Q^2$  one can get $G_M^s (Q^2 )$ with some
contamination from the axial part. This is the kinematical regime where the
SAMPLE experiment [54] hopes to determine the strange anomalous magnetic
moment. This warrants a closer look. In a simplified analysis, we set $s^2 =
1/4$, $\theta = \pi$ and work at some low $Q^2$, say 0.2 GeV$^2$. Then,
$A(Q^2)$ takes the simple form [55]
$$A(Q^2) \simeq -{G_F Q^2 \over \sqrt{2} \pi \alpha} {G_M^Z \over G_M^\gamma}
\simeq -{G_F Q^2 \over 4 \sqrt{2} \pi \alpha} {\kappa_p - F^0_2 (0) \over
\kappa_p + 1} = -{G_F Q^2 \over 4 \sqrt{2} \pi \alpha} \, R \eqno(3.22)$$
assuming that the $Q^2$-dependence is similar for $G_M^Z$ and $G_M^\gamma$.
The value of $R$ ranges form $R=1 $ to $R=0.28$ for $F_2^0 (0) = -1 \ldots 1$.
For no strangeness contribution, we have $F_2^0 (0) = -0.12$ and thus $R =
0.64$. To give an idea about what one can expect, let me consider some models.
The $SU(3)$ Skyrme model with a proper treatment of the symmetry breaking
leads to $F_2^0 (0) = -0.17 \ldots -0.25$ [56], the NJL model to $F_2^0 (0) =
-0.16 \ldots -0.26$ [57] and vector meson dominance  supplemented with the
$\phi$-meson (i.e. explicit strange quarks) to $F_2^0 (0) = -0.43 \pm 0.09$
[58]. Similarly, one can also discuss the strange electric radius, i.e. the
radius related to the operator $\bar s \gamma_\mu s$.  A simple model based on
a kaon cloud surrounding the nucleon leads to processes like $ p \to \Lambda
K^+, \, \Sigma^0 K^+, \ldots \to p$ and lets us expect that $<r^2_E>_s$ is
positive since the positive charge sits in the cloud of the kaons. This is,
however, a very crude estimate. In fact, most models tend to give a small and
negative $<r^2_E>_s$, like e.g. VMD plus the $\phi$-meson $-0.14 \pm 0.04$
fm$^2$ [58] or the $SU(3)$ Skyrmion $-0.10 \pm 0.05$ fm$^2$ [56]. These are
fairly small numbers and it appears doubtful that the strange electric radius
can be determined accurately from a proton target in the near future
(assuming, of course, that its value is as small as indicated by the models).
Here, nuclei might come in handy.  For isoscalar, spin-0 nuclei such as $^4$He
or $^{12}$C, one finds for the asymmetry [59]
$$A =  {G_F Q^2 \over 4 \sqrt{2} \pi \alpha} \bigl[ s^2 + {G_E^s \over 2 (
G_E^p + G_E^n )} \bigr] \, \, , \eqno(3.23)$$
which at very low $Q^2$ allows to extract the weak mixing angle and at
somewhat higher momenta the strange electric ff. Of course, one has to take
into account nuclear structure issues like pv level mixing and so on. A
lucid discussion of these topics can be found in refs.[59,60]. Clearly, pv
electron scattering is an interesting field and many more aspects of it not
covered here can e.g. be found in the proceedings of the CALTECH workshop
[61].
\titleb{3.5}{Neutrino and
Antineutrino Scattering off Nucleons and Nuclei}
Neutrino scattering off nucleons or nuclei offers another possibility of
exploring the strange quark content of the nucleon. The neutrino couples via
the $Z^0$ to the quarks and thus probes isosinglet, triplet and octet
components. Let us first consider the differential cross section for elastic
neutrino/antineutrino-proton scattering,
$${d \sigma \over d Q^2} = {G_F^2 m^2 \over 8 \pi E_\nu^2} \bigl[ A \pm B \, W
+ C \, W^2 \bigr] \, \, , \eqno(3.24)$$
with $E_\nu$ the neutrino energy, $W = (4 E_\nu m - Q^2 )/m^2$ and the $'\pm
'$ refers to the case of neutrinos (antineutrinos). The functions $A$, $B$ and
$C$ depend on the ffs $G_1$ and $F_{1,2}^Z$ (I omit the superscript '$Z$' on
$F_1$ and $F_2$),
$$\eqalign{A &=4 \tau \bigl[ G_1^2 (1 + \tau ) - 4 (1 - \tau ) (F_1^2 - \tau
F_2^2 ) + 16 \tau F_1 F_2 \bigr] \, \, , \cr
B &= - 8 \tau G_1 ( F_1 + F_2 ) \, \, , \cr
C &= G_1^2 / 4 + F_1^2 + \tau F_2^2 \, \, . \cr} \eqno(3.24a)$$
To demonstrate this in more detail, let me simplify the analysis by assuming a
$Q^2$ value of about 0.5 GeV$^2$ and retaining only the terms linear in the
strange ffs. This allows to recast (3.24) in the form [62]
$${d \sigma \over d Q^2} = \sigma_{{\rm non-strange}} \bigl[ 1 - 0.72 F_1^s
(Q^2 ) - 0.78 F_2^s (Q^2 ) - 1.23 G_1^s (Q^2 ) \bigr] \, \, . \eqno(3.24b)$$
Obviously, to extract the strange matrix element $<p|{\bar s} \gamma_\mu
\gamma_5 s |p> = G_1^s (0)$, one has to know something about $F^s_{1,2} (Q^2
)$ or make some assumptions. Ahrens et al. [63] and Kaplan and Manohar [50]
have shown that setting $F_1^s = F_2^s = 0$ and assuming that all three axial
ffs have the same dipole mass ($M_A = 1.032$ GeV), the extracted value of
$G_1^s (0)$ is in agreement with the one obtained from polarized deep
inelastic $\mu p$ scattering [36]. However, as already stressed in ref.[63],
this result is very sensitive to the actual value of the diple mass $M_A$.
This issue was further addressed by Bernard et al. [64]. They used the
topological chiral soliton model of the nucleon to calculate the isosinglet
axial ff since in the framework of this model the known triplet ff is well
described. It was found that the cut-off mass in the singlet channel is
actually 20 per cent larger than the one for the triplet and octet ffs.
Redoing the analysis with these ffs, one finds that the value of $G_1^s (0)$
is reduced by a factor of  three. A similar analysis has recently been
performed by Garvey et al.[65]. In conclusion, the often claimed agreement
concerning the value of $G_1^s (0)$ from the neutrino and EMC data can only be
considered accidental. To avoid the problems of extrapolating from the typical
$Q^2$ of 0.5 to 1 GeV$^2$ in elastic $\nu p$ or ${\bar \nu}p$ scattering, it
was recently proposed that (anti)neutrino induced quasi-free nucleon knock-out
off nuclei might give a better handle on the strange  matrix elements [66]. In
the quasi-free region, one  has small momentum transfers and typical neutrino
energies of 0.2 GeV. The ratio of the proton to the neutron yield depends
sensitively on $G_1^s$ and $F_2^s$,
$$R = Y_p \, / \, Y_n = F ( G_1^s, \, F_2^s ) \,\, . \eqno(3.25)$$
For example, if $F_2^s (0) = -0.22$, $R_{\bar \nu}$ varies between 0.85 and
1.9 for $G_1^s (0)$ between 0 and -0.2. For further details, see ref.[66]. Of
course, it is mandatory to understand well the nuclear structure issues, in
particular one assumes that the knock-out is involving only one nucleon. How
good these assumptions are is still under debate. What is important is to
realize that neutrino and pv electron scattering processes nicely complement
each other and that many different experiments have to be performed to
ultimately pin down the strength of the strange matrix elements in the proton.
\titleb{3.6}{Down and Dirty: QED, QCD and Heavy Quark Corrections}
Up to now, I have entirely worked at tree level. At low energies, radiative
corrections and effects from heavy quarks are expected to be small. However, in
certain cases the tree level couplings are suppressed or vanish (as it is the
case with the isocalar axial current) or one is trying to extract small
numbers. Therefore, it is mandatory to investigate the effects of QED and QCD
corrections on the processes considered before. Since the neutrino has no
direct QED and QCD couplings, let me start with the discussion of corrections
to neutrino-hadron (quark) scattering. To be specific, consider the induced
isoscalar coupling. At tree level it vanishes, but that does not mean that it
is
zero alltogether. The basic Feynman diagram which induces such a coupling is
the famous triangle diagram. On one end, the axial part ($\gamma_\mu \gamma_5$)
of the neutrino current couples and the other two corners are attached via
gluon exchanges to the external light (u,d,s) quark. In the loop, quarks of all
flavors run around. A simple one-gluon exchange is forbidden by color
neutrality. This is the basic diagram to lowest order in the strong
interactions. The contribution of such a type of diagram has to be finite since
the sum over all quark flavors leads to anomaly cancellation,
$$ \sum_i T_3^i = 0 \, \, , \eqno(3.26)$$
where $T_3$ denotes the weak isospin. However, we are interested at scales much
below the intermediate vector boson and heavy quark masses. In this energy
regime, the triangle diagram leads to an induced isoscalar current which is not
suppressed by inverse powers of the heavy quark masses as naive decoupling
would suggest. Indeed, a straightforward perturbative calculation of this
two-loop diagram yields [67]
$$A_\mu^Z = {1 \over 2} (A_u - A_d ) + (A_u + A_d ) {1 \over 4} \bigl({ g^2
\over 4 \pi} \bigr)^2 \ln {\Pi_i m^2_{i+} \over \Pi_i m^2_{i-}} \, ,
\eqno(3.27)$$
where $m_{i \pm}$ refers to the masses of the quarks with weak isospin $\pm
1/2$, i.e. with charges 2/3 and -1/3, respectively, and $g$ is the strong
coupling constant. Notice that the
induced isoscalar current $(A_u + A_d )$
depends logarithmically on the quark masses and not on inverse powers of them
(as advertised). Collins, Wilczek and Zee [67] have shown how to do this in a
more systematic fashion. Their method is based on integrating out quark
doublets
in succession. Denote by $H = A_t - A_b$ the axial current of the heaviest
doublet. The idea is to recast $H$ in the form
$$H = \sum_i A_i {\bar L}_i + \sum_i B_i {\bar H}_i \, \, , \eqno(3.28)$$
where the light (${\bar L}_i$) and the heavy (${\bar H}_i$) particle operators
exhibit decoupling. This means that the matrix elements of the ${\bar L}_i$ see
the heavy quarks only via power law corrections. Repeating this exercise for
$\tilde{H} = A_c - A_s$, the induced isoscalar current reads
$$\Delta A_\mu^Z = (A_u + A_d) {1 \over 4} {\alpha_s (m_s) \over \pi}
{\alpha_c (m_c) \over \pi} \ln {m_c^2 \over m_s^2} \simeq 0.05
\, (A_u + A_d) \, \, ,
\eqno(3.29)$$
where I have neglected the much smaller contribution from the $(t,b)$ doublet.
Of course, this approach is too bold since at the scale of the strange quark
mass $\lambda = m_s \simeq 175$ MeV, this perturbative treatment can not be
justified any more. Therefore, Kaplan and Manohar [50] have generalized the
method of ref.[67] to integrate out the various quarks separately. Their
argument is based on the assumption $M_Z > m_t$. This simplifies the analysis
but is not mandatory. In this case, integrating out the $Z^0$ one has an
effective Lagrangian for neutrino-quark scattering,
$${\cal L}_{\rm eff} = - {G_F \over \sqrt{2}} \sum_{u,d,s,c,b,t} {\bar \nu}
\gamma^\mu (1 - \gamma_5 ) \nu \bigl[ {\bar q} (T_3 - 2 s^2 Q) \gamma_\mu q -
{\bar q} T_3 \gamma_\mu \gamma_5 q \bigr] \, \, . \eqno(3.30)$$
In principle, one should now calculate corrections to this effective
Lagrangian. However, due to the absence of direct QED or QCD couplings of the
neutrino, one can rewrite these corrections as corrections to the hadronic
current. The procedure goes as follows.  One uses the renormalization group to
evolve (3.30) down to the scale of the top mass. At that energy, one integrates
out the t quark. This induces two types of corrections, one class scaling like
$g^2 (m_t) / m_t^2$ (i.e. exhibiting decoupling). Since the t is removed from
the theory, the $Z^0$ axial current has an anomaly because (3.26) is not
fulfilled any more. This leads to a multiplicative renormalization of the
singlet axial current, which is the most important effect as we will see. One
then proceeds by scaling down to $\lambda = m_b$, integrating out the b and so
on until one reaches the scale $\lambda = 1$ GeV (which is a typical energy in
the elastic $\nu p$ scattering process and large enough to justify the
perturbative treatment). Putting the important pieces from the
multiplicative renormalizations together, the relation
between the axial current at $\lambda = M_Z$ and at $\lambda = 1$ GeV reads
[50]
$$\eqalign{A_\mu^Z (\lambda = 1 \, {\rm GeV}) &= A_\mu^Z (\lambda = M_Z ) + {1
\over 2} (A_u + A_d) \Delta_A  \cr
\Delta_A &=  \Lambda_{\mu c} \biggl[ {1 \over 2} + {3 \over 10} \Lambda_{cb}
\bigl( \Lambda_{bt} - 1 \bigr) \biggr]               \cr
\log \Lambda (m_i , m_j , N_f ) &= N_f {\alpha_s (m_i ) \over \pi}
{\alpha_s (m_j ) \over \pi} \log {m_i^2 \over m_j^2}\, \, , \cr} \eqno(3.31)$$
which leads to $\Delta A = 0.02$ [50]. Obviously one recovers the result (3.29)
if one were to integrate out the s quark too. Notice that the finite pieces due
to the renormalization group evolution are negligible and that there is also
an induced vector current with $\Delta_V < 10^{-4}$. The main effect of the
photons comes from the penguin diagram leading to a running of the weak mixing
angle [68]
$$ s^2 (\lambda^2 ) = s^2 (M_W^2 ) + \sum_j {\alpha \over 3 \pi} \ln \biggl(
{M_W^2 \over \lambda^2} \biggr) Q_j ( T_{3j} - 2 s^2 Q_j ) \, \, ,
\eqno(3.32)$$
where the index $j$ extends over the quark flavours which are integrated out.
In addition, there are pure next-to-leading order electroweak effects. These
have been considered in ref.[69]. The radiatively induced isoscalar axial
current has a very small coefficient,
$$-{3 \alpha \over 16 \pi s^2} \biggl[ 1 + {1 \over 2 c^2} \bigl( 1 - 2 s^2 +
{20 \over 9} s^4 \bigr) \biggr] = - 0.003 \,\, , \eqno(3.33)$$
which is negligible compared to the heavy quark effects. Therefore, if one
restrcits oneself to scales above $\lambda \simeq 1$ GeV, the QED and QCD
corrections to neutrino-hadron scattering are known and under control. Matters
become more messy if one attempts to go to lower energies and also if one
considers electron-hadron scattering as I will do now. In this case, we have in
addition the direct photon exchange which gives rise to further complications.
What we are after can be summarized as follows. Denote by $C$ the tree level
value of any coupling between the lepton and the hadron (quark) currents. Due
to the radiative corrections, these couplings are modified
$$ C = C [ 1 + R ] \, \, , \eqno(3.34)$$
so that $R$ gives the ratio of the corresponding value including radiative
corrections to the tree level one (in case that $C$ happens to vanish at tree
level, one has to choose another coupling as the reference point). Naively, one
expects the size of these corrections to be small, the typical scale being
$$ G_F m^2 {\alpha \over 4 \pi} \le 10^{-8} \, \, . \eqno(3.35)$$
This rather small number is sometimes considerably enhanced and, furthermore,
the radiative corrections can in some circumstances compete with the effects
induced by the strange quarks. Holstein and Musolf [70] have given the most
detailed evaluation of these effects. Clearly, the results of their
calculations should be considered indicative since considerable uncertainties
are involved as will be discussed below. A very basic and introductory
presentation concerning the calculation of radiative corrections for the pv
processes under consideration has been given by Musolf [71] which the reader
not familiar with the concepts of renormalization in the standard
model certainly will appreciate. In essence, there are two classes of Feynman
diagrams contributing at one loop order.  The first class involves only one
quark in the nucleon. Typical representatives are vertex or vector
boson propagator corrections.  More complicated are the diagrams involving two
quarks, such as the $\gamma Z^0$ box or exchange current type diagrams where a
massive vector bosons is exchanged between two quarks while the photon couples
to one of them (many pictorials are displayed in ref.[71]). Let me first
consider the one quark type diagrams. These depend on the not yet known masses
of the t quark and the Higgs boson and, if there is physics beyond the
standard model, implicetely on the parameters of this new physics. This latter
dependence is parametrized in terms of S, T and U or $\epsilon_1$,
$\epsilon_2$ and $\epsilon_3$ (as discussed by Altarelli here [49]). There are
two effects which can conspire to give much larger values for these loop
corrections than the estimate (3.35) would suggest. These are large logarithms
of the light fermion to the vector boson mass ratios and the suppression of
tree level couplings. A particularly illustrative example is the scattering
process $\nu_e \mu \to \nu_e \mu$. While the tree level $Z^0 - \mu$ coupling
is suppressed, the $W$-vertex correction to the $Z^0-\nu_e$ coupling induces
a log($m_e^2 / M_W^2 $), with $m_e$ the electron mass. The corresponding ratio
follows to be [70]
$$R(\nu_e \mu \to \nu_e \mu ) = {\alpha \over 4 \pi} {8 \over 3} {\ln (m_e^2 /
M_W^2 ) \over 4 s^2 - 1} \simeq 0.5 \, \, . \eqno(3.36)$$
This calculation is clean in that it involves only leptons. For  the case of
electron-hadron scattering, the quarks are bound within a nucleon of a size
of about 1 fm. Using the uncertainty principle, this allows one to get an idea
about the typical quark momenta. However, this is a very crude estimate and
induces some uncertainty (for details, see ref.[70]). Furthermore, it is not
obvious which quark mass values one should insert. When one considers quark
loop corrections to the vector boson propagators, one obviously deals with the
current quarks. In contrast, in the various box diagrams it is more justified
to use the so-called constituent masses ( $\sim$ 330 MeV for the u and
d quarks) to account for the binding effects. These subtleties are also
discussed in refs.[70,71]. For processes of the type $V(e) \times A(N)$ (which
means that the vector current of the electron couples to the nucleonic axial
current), one finds corrections $R^p = -0.65 \ldots -0.28$ and
$R^n = -0.56 \ldots -0.06$ for the proton and the neutron, in order. These are
clearly larger than the dimensional argument (3.35) suggests. Furthermore, we
also have to consider the many quark diagrams. Here, the situation is less
transparent, i.e. it is much more difficult to deal with this class. To
proceed, one can resort to some kind of "hadronic duality" [70,71,72].
This means that one considers a meson cloud picture instead of the more
complex quark diagrams. As examples, the excitation of quark-antiquark pairs
with the quantum numbers of a pion probed by the photon and with a successive
vector boson exchange translates to the photon coupling to a pion in flight,
with one pion-nucleon coupling parity-conserving (strong) and the other pv
(weak). Similarly, other diagrams      translate into a $\gamma \rho^0$
conversion followed by a pv $\rho N$ coupling. Such diagrams contribute to the
nucleon axial current via [70,72]
$$\eqalign{ \delta A_N^\pi &= {g_{\pi N} h_{\pi N} \over m^2} \biggl[ {\pi
\over 6}{m \over M_\pi} + \ln \bigl( {M_\pi \over m} \bigr) + \ldots \biggr]
\, \, , \cr \delta A_N^\rho &= {h_{\rho N} \over g_{\rho N}}{ 1 \over
m^2_\rho} \, \, . \cr} \eqno(3.37)$$
The strong coupling constants ($g_{\pi N}, \, g_{\rho N}$) are fairly well
known. Matters are different for the parity-violating couplings
($h_{\pi N}, \, h_{\rho N}$). Their calculation has been pioneered by Donoghue
et al.[73], a recent update including additional experimental constraints and
chiral soliton model calculations can be found in refs.[74.75]. The pion cloud
contribution diverges in the chiral limit which is no surprise in the light of
the discussion presented in section 2. Ultimately, CHPT methods might shed
some light on these long-distance contributions. In ref.[70], the estimates
for processes of the type $V(e) \times A(N)$ range from  $-0.07 \ldots +0.37$
for $R^p$ and $-0.07 \ldots +0.24$ for $R^n$. Combining these numbers with the
ones from the one quark diagrams, one sees that there are potentially large
corrections. To make all this more transparent, let us consider again the
determination of the strange anomalous magnetic moment from pv electron-proton
scattering. In backward direction and at small $Q^2$ we have [76]
$$A_{ep} \simeq -{G_F Q^2 \over 4 \sqrt{2} \pi \alpha} \biggl\lbrace {\kappa_p
\over \mu_p}\bigl[ 1 + R^M_{{\rm strange}} + R^M_{{\rm rad}} \bigr] - {2 m (
E' + E )\over Q^2} {1 - 4 s^2 \over \mu_p} g_A ( 1 + R^A_p ) \biggr\rbrace \,
\, , \eqno(3.38)$$
where $R^M_{{\rm strange, \, rad}}$ denotes the strangeness  and radiative
corrections to the magnetic moment and $R^A_p$ the one to the proton axial
current. Two important observations are in order. First, the radiative
corrections to the magnetic asymmetry are comparable to the strange quark
induced ones,
$$ {R^M_{{\rm rad}} \over R^M_{{\rm strange}} } \simeq { 0.1 \over \mu_s} \,
\, . \eqno(3.39)$$
Second, there is a large correction to the background  of the axial asymmetry
which makes up roughly 30 percent of the signal for the SAMPLE kinematics.
These two effects severely constrain the accuracy for determining the strange
anomalous magnetic moment. To overcome these problems, it is therefore
mandatory to perform many complementary experiments like e.g. backward-angle
${\vec e}d$ scattering, pv quasi-elastic scattering and so on. For a nice
overview about these topics, I refer to ref.[76]. Finally, let me stress again
that more theoretical effort is needed to further tighten the limits on these
radiative corrections so that an unambigous interpretation of the experiments
will be possible.

\titlea{4}{Summary and Outlook}

The standard model of the strong and electroweak interactions is an
embarrasingly successful theory.  It is least understood at its extreme energy
limits. First, at very high energies, it is not known what exactly triggers
the spontaneoaus symmetry breaking of $SU(2)_L \times U(1)_Y \to U(1)_{{\rm
em}}$ at a scale of $< \phi > = 250$ GeV. While the standard Higgs boson does
the job, one is left with just too many free parameters to feel confident with
such a scenario. At present, the physics beyond the standard model hides
itself quite effectively as discussed by Altarelli at this school [49]. Future
high-energy colliders like the LHC or the SSC  will hopefully shed light on
the electroweak symmetry breaking sector. May be less spectacular, but as
challenging is the problem of hadron structure in the non-perturbative regime.
Chiral perturbation theory is a method which allows to work out systematically
the consequences of the spontaneous chiral symmetry breaking in QCD. It is
based on a simultaneous expansion in the (small) external momenta  and (light)
quark masses (u,d,s) and enjoys considerable success in the meson sector. In
the first part of these lectures, I have mostly been concerned with the
problems surrounding the inclusion of matter fields in the chiral expansion.
While this is technically a straightforward procedure based on chiral counting
rules, it is conceptually less transparent than in the meson sector due to the
appearance of the nucleon mass term. This is a scale of the order of the
chiral symmetry breaking scale also it is not related to  the symmetry
violation and, furthermore, it does not vanish in the chiral limit. I have
concentrated here on a few selected processes in the two-flavor sector like
nucleon Compton scattering or pion photo- and electroproduction. Since the u
and d quarks are really light, the corresponding expansion parameters $M_\pi /
4 \pi F_\pi$ and $E_{{\rm pion}} / F_\pi$ are small and one has a better
change  of a converging chiral expansion. However, as we have seen in the
discussion of the LET (2.14), the appearance of the nucleon mass renders these
dimensional arguments rather dangerous. It appears at this point that in the
heavy mass formulation of baryon CHPT such terms should not show up since the
baryon propagator does not include the nucleon mass. However, in that case the
large new scale is hidden by means of the Goldberger-Treiman relation in large
prefactors accompanying the axial-vector coupling strength $g_A$.  All
calculations performed so far have been carried out in the one-loop
approximation indicating that it is mandatory to include at least the terms of
order $q^4$ (which are beyond next-to-leading order). This argument is further
strengthened when one considers the three-flavor sector. Here, one very often
finds large kaon and eta loop corrections which make one feel uneasy about the
validity of the chiral expansion [77]. The suggestion that these large loop
effects are largely cancelled by contributions from the spin-3/2 decuplet in
the intermediate states [16,18] is appealing but has not yet been put on a
firm basis, which means that one has to  perform a full order $q^4$  (or
higher)
calculation. There are, however, on-going activities in this area and further
progress can be expected soon.

The second topic of these lectures concerned the quark structure of the
nucleon at low energies. I have discussed how particular combinations of the
electromagnetic and weak form factors, which parametrize the structure of the
nucleon and its flavor decomposition, allow to extract interesting matrix
elements like the strange anomalous magnetic moment, $F_2^s (0)$, or the
strange axial nucleon charge, $G_1^s (0)$. Parity-violating electron
scattering
($\gamma Z^0$-interference) and (anti)neutrino scattering off nucleons or
nuclei are the tools to determine these and other matrix elements like e.g.
the extension of the nucleon as given by the strange electric radius. However,
and that was a major topic here, since in most cases one tries to extract
small numbers, one also has to account for radiative and heavy quark
corrections. While these are nominally small, they often tend to be enhanced
by large factors way above the naive dimensional estimates. In the most
extreme case (cf. the isoscalar axial coupling), the tree level couplings with
which one works at leading order are vanishing. In neutrino-hadron scattering,
if one does not go below a typical scale of say 1 GeV, these various
corrections are under control. As a particular example, I have discussed the
induced isoscalar axial coupling, which is not only empirically interesting
but also shows how under certain circumstances effective field theory methods
can and {\it should} be used to caluclate QED, QCD and heavy quark corrections
[50]. When it comes to electron-hadron scattering, the situation is much less
satisfactory. First, as I discussed, in many experiments one wants to work at
much smaller energy scales and therefore has to account for the effects of
quark confinement. Second, the direct photon coupling to the hadrons (quarks)
gives rise too much more diagrams than it is the case in neutrino scattering.
The most systematic analysis of radiative corrections to parity-violating
electron-hadron scattering [70] translates the Feynman diagrams involving more
than one quark in the nucleon into "hadronic ones", i.e. making use of the
meson cloud picture. This induces uncertainties which are of the order of the
calculated effects themselves. Another source of uncertainty is the momentum
distribution of the bound quarks which one can only account for in an
approximate fashion. Obviously, the theorists are called for providing better
calculations of this type so that the experimenters have a better change to
extract the various strange matrix elements. Further complications arise when
one considers the scattering off nuclei [78]. As I pointed out, many
complimentary experiments which are sensitive to different combinations of the
form factors and have different next-to-leading order corrections have to and
will be performed at MAMI, CEBAF, $\ldots$ [79]. These experiments are eagerly
awaited. To come back to the first part of the lectures, once we will have
learned how to treat the three-flavor sector of baryon CHPT, one will also be
able to discuss the strange matrix elements and confront the theoretical
predictions with the data.

\titlea{5}{Acknowledgements}

I would like to thank the organizers, in particular Professor W. Plessas, for
their invitation and kind hospitality extended to me. This work was supported
in part by the Deutsche Forschungsgemeinschaft through a Heisenberg
fellowship and the Schweizerischer Nationalfonds.
\vskip 1truecm
\begref{References}{[MT1]}
\refno{[1]}
J. Goldstone, {\it Nuovo Cim.\/}
{\bf 19} (1961) 154
\refno{[2]}
H. Leutwyler, in ``Recent Aspects of Quantum Fields'', eds. H. Mitter
and M. Gausterer, Springer Verlag, Berlin, 1991
\refno{[3]}
Ulf-G. Mei{\ss}ner, Recent Developments in Chiral Perturbation
Theory, preprint BUTP-93/01 (1993)
\refno{[4]}
S. Weinberg, {\it Phys. Rev. Lett.\/} {\bf 17} (1966) 616
\refno{[5]}H. Lehmann, {\it Phys. Lett.\/} {\bf
B41} (1972) 529; {\it Acta Phys. Austriaca Suppl.\/} {\bf 11} (1973) 139
\refno{[6]}S. Weinberg, {\it Physica} {\bf 96A} (1979) 327
\refno{[7]}J. Gasser and H. Leutwyler, {\it Ann. Phys. (N.Y.)\/}
 {\bf 158} (1984) 142
\refno{[8]}G. Ecker, J. Gasser, A. Pich and E. de Rafael,
{\it Nucl. Phys.\/} {\bf B321}
(1989) 311;
J. F. Donoghue, C. Ramirez and G. Valencia,
{\it Phys. Rev.\/} {\bf D39}
(1989) 1947
\refno{[9]}J. Gasser and Ulf-G. Mei{\ss}ner,
{\it Phys. Lett.\/} {\bf B258}
(1991) 219
\refno{[10]}J. Gasser and Ulf-G. Mei{\ss}ner,
{\it Nucl. Phys.\/} {\bf B357} (1991) 90
\refno{[11]}S. Coleman, J. Wess and B. Zumino,
{\it Phys. Rev.\/} {\bf 177} (1969) 2239;
C. G. Callan, S. Coleman, J. Wess and B. Zumino,
{\it Phys. Rev.\/} {\bf 177} (1969) 2247
\refno{[12]}A. Krause, {\it Helv. Phys. Acta\/} {\bf
63} (1990) 3
\refno{[13]}J. Gasser, M.E. Sainio and A. ${\check {\rm S}}$varc,
{\it Nucl. Phys.\/}
 {\bf
B307} (1988) 779
\refno{[14]}E. Jenkins and A.V. Manohar, {\it Phys. Lett.\/} {\bf B255} (1991)
558
\refno{[15]}S. Weinberg, {\it Nucl. Phys.\/} {\bf
B363} (1991) 3
\refno{[16]}E. Jenkins and A.V. Manohar, in "Effective Field Theories of the
Standard Model", ed. Ulf--G. Mei{\ss}ner, World Scientific, Singapore,
1992
\refno{[17]}
Ulf-G. Mei{\ss}ner,
{\it Int. J. Mod. Phys.}
{\bf E1} (1992) 561
\refno{[18]}E. Jenkins and A.V. Manohar, {\it Phys. Lett.\/} {\bf B259} (1991)
353
\refno{[19]} V. Bernard, N. Kaiser and Ulf-G. Mei{\ss}ner,
in preparation
\refno{[20]}
N.M. Kroll and M.A. Ruderman,
{\it Phys. Rev.\/} {\bf 93} (1954) 233
\refno{[21]}
V. Bernard, J. Gasser, N. Kaiser and Ulf-G. Mei{\ss}ner,
{\it Phys. Lett.\/} {\bf B268} (1991) 291
\refno{[22]}
V. Bernard, N. Kaiser and Ulf-G. Mei{\ss}ner,
{\it Nucl. Phys.\/}
{\bf B383} (1992) 442
\refno{[23]}A. M.
Bernstein and B.R. Holstein, {\it Comments Nucl. Part. Phys.\/} {\bf 20}
(1991) 197
\refno{[24]}
D. Drechsel and L. Tiator,
{\it J. Phys. G: Nucl. Part. Phys.\/} {\bf 18} (1992) 449
\refno{[25]}
V. Bernard, N. Kaiser and Ulf-G. Mei{\ss}ner,
{\it Phys. Lett.\/} {\bf B282} (1992) 448
\refno{[26]}J. H. Koch, S. Scherer and J. L. Friar, to be published, and
S. Scherer and J. H. Koch,
{\it Nucl. Phys.\/} {\bf A534}
(1991) 461
\refno{[27]}
T. P. Welch et al., {\it Phys. Rev. Lett.\/}
{\bf 69} (1992) 2761
\refno{[28]}
V. Bernard, N. Kaiser, T.--S. H. Lee and Ulf-G. Mei{\ss}ner,
{\it Phys. Rev. Lett.} {\bf 70} (1993) 387
\refno{[29]}
Y. Nambu and D. Luri\'e, {\it Phys. Rev.\/} {\bf 125}
(1962) 1429;
Y. Nambu and E. Shrauner, {\it Phys. Rev.\/} {\bf 128}
(1962) 862
\refno{[30]}
V. Bernard, N. Kaiser and Ulf-G. Mei{\ss}ner,
{\it Phys. Rev. Lett.\/} {\bf 69} (1992) 1877
\refno{[31]}F.J. Federspiel et al., {\it Phys. Rev. Lett.\/} {\bf 67} (1991)
1511
\refno{[32]} F.E. Low, {\it Phys.  Rev.\/} {\bf 96} (1954) 1428;
M. Gell-Mann and M.L. Goldberger, {\it Phys. Rev.\/} {\bf 96} (1954) 1433
\refno{[33]}
V. Bernard, N. Kaiser, J. Kambor and Ulf-G. Mei{\ss}ner, {\it Nucl. Phys.\/}
{\bf B388} (1992) 315
\refno{[34]}R.L. Workman and R.A. Arndt, {\it Phys. Rev.\/} {\bf D45} (1992)
1789
\refno{[35]}S.D. Drell and A.C. Hearn, {\it Phys. Rev. Lett.\/} {\bf 16}
(1966) 908;
S.B. Gerasimov, {\it Sov. J. Nucl. Phys.\/} {\bf 2} (1966) 430.
\refno{[36]}J. Ashman {\it et al.}, {\it Phys. Lett.\/} {\bf B206} (1988) 364.
\refno{[37]}
V. Bernard, N. Kaiser and Ulf-G. Mei{\ss}ner, Bern University preprint
BUTP-92/51, 1992
\refno{[38]}
V. Burkert and Z. Li, {\it Phys. Rev.} {\bf D47} (1993) 46
\refno{[39]}
G. H\"ohler, in Landolt--B\"ornstein, vol.9 b2, ed. H. Schopper,
 Springer, Berlin, 1983
\refno{[40]}H. Arenh\"ovel, these proceedings; B. Schoch, these proceedings
\refno{[41]}R.E. Taylor, these proceedings
\refno{[42]}S. Platchkov et al., {\it Nucl.
Phys.\/} {\bf A508} (1990) 343
\refno{[43]}
M. Gari and W. Kr\"umpelmann, {Phys. Lett.\/} {\bf B274} (1992) 159;
A. Lung et al.,
{\it Phys. Rev. Lett.\/} {\bf 70} (1993) 718
\refno{[44]}
S.J. Brodsky and G.P. Lepage, {\it Phys. Rev.\/} {\bf D22} (1980) 2157
and references therein
\refno{[45]}
Ulf-G. Mei{\ss}ner, {\it Phys. Rep.\/}
{\bf 161} (1988) 213
\refno{[46]}D. Beck, {\it Phys. Rev.\/} {\bf D39} (1989) 3248
\refno{[47]}R. Cahn and F. Gilman, {\it Phys. Rev.\/} {\bf D17} (1978) 1313
\refno{[48]}D. Beck, University of Illinois preprint NPL-91-015, 1991
\refno{[49]}G. Altarelli, these proceedings
\refno{[50]}D. Kaplan and A.V. Manohar, {\it Nucl.
Phys.\/} {\bf B310} (1988) 527
\refno{[51]}E. Reya, these proceedings
\refno{[52]}
J. Gasser, H. Leutwyler and M.E. Sainio, {\it Phys. Lett.\/}
 {\bf 253B} (1991) 252, 260
\refno{[53]}C.Y. Prescott {\it et al.}, {\it Phys. Lett.\/} {\bf B77}
(1978) 1347; {\it Phys. Lett.\/} {\bf B84} (1979) 524
\refno{[54]}BATES experiment 89-06, see e.g. D. Beck in ref.[61]
\refno{[55]}R.D. McKeown, {\it Phys. Lett.\/} {\bf B219} (1989) 40
\refno{[56]}N.W. Park and H. Weigel, {\it Nucl. Phys.\/} {\bf A541}
(1992) 453
\refno{[57]}V. Bernard and Ulf-G. Mei{\ss}ner, {\it Phys. Lett.\/} {\bf B216}
(1989) 392; {\bf B223} (1989) 439
\refno{[58]}R.L. Jaffe, {\it Phys. Lett.\/} {\bf B229} (1989) 275
\refno{[59]}M.J. Musolf and T.W. Donnelly, {\it Nucl. Phys.\/} {\bf A546}
(1992) 509
\refno{[60]}T.W. Donnelly et al., {\it Nucl. Phys.\/} {\bf A541}
(1992) 525
\refno{[61]}"Parity Violation in Electron Scattering", eds. E. Beise and R.
McKeown, World Scientific Publ., Singapore, 1990
\refno{[62]}E. Beise and R. McKeown, {\it Comm. Nucl. Part.
Phys.\/} {\bf 20} (1991) 105
\refno{[63]}L.A. Ahrens et al., {\it Phys. Rev.\/} {\bf D35} (1987) 785
\refno{[64]}
V. Bernard, N. Kaiser and Ulf-G. Mei{\ss}ner, {\it Phys. Lett.\/}
{\bf B237} (1990) 545
\refno{[65]}G.T. Garvey, W.C. Louis and D.H. White, Los Alamos preprint
LA-UR-93-0037, 1992
\refno{[66]}G.T. Garvey et al., {\it Phys. Lett.\/} {\bf B289} (1992) 249
\refno{[67]}J. Collins, F. Wilczek and A. Zee, {\it Phys. Rev.} {\bf D18}
(1978) 242
\refno{[68]}
H. Georgi, ``Weak Interactions and Modern Particle Physics'',
Benjamin/Cummings, Reading, MA, 1984
\refno{[69]}R.N. Mohapatra and G. Senjanovic, {\it Phys. Rev.} {\bf D19} (1979)
2165
\refno{[70]}M.J. Musolf and B.R. Holstein, {\it Phys. Lett.\/} {\bf B242}
(1990) 461
\refno{[71]}M.J. Musolf in ref.[61]
\refno{[72]}W.C.Haxton, E.M.Henley and M.J. Musolf, {\it Phys. Rev. Lett.\/}
{\bf 63} (1989) 949
\refno{[73]}B. Desplanques, J.F. Donoghue and B.R. Holstein, {\it Ann. Phys.
\/} {\bf 124} (1980) 449
\refno{[74]}
N. Kaiser and Ulf-G. Mei{\ss}ner, {\it Nucl. Phys.\/}
{\bf A499} (1989) 699
\refno{[75]}
Ulf-G. Mei{\ss}ner, {\it Mod. Phys. Lett.\/}
{\bf A5} (1990) 1703
\refno{[76]}M.J. Musolf, MIT preprint CTP 2120, 1992
\refno{[77]}J. Bijnens, H. Sonoda and M.B. Wise, {\it Nucl.Phys.\/} {\bf B261}
(1985) 185
\refno{[78]}C.J. Horowitz and J. Piekarewicz, Florida State University
preprint, FSU-SCRI-93-18, 1993
\refno{[79]}Some examples are MAMI-A4, CEBAF 91-004, CEBAF 91-010 and CEBAF
91-017
\endref
\byebye

